\documentclass[%
 reprint,
superscriptaddress,
 amsmath,amssymb,
 aps,
]{revtex4-2}

\usepackage{soul}
\usepackage{tikz}
\usetikzlibrary{shapes.geometric, arrows}
\usepackage{graphicx}
\usepackage{dcolumn}
\usepackage{bm}
\usepackage{hyperref}

\usepackage[version=4]{mhchem}
\usepackage{siunitx}
\usepackage{comment}
\usepackage{tabularx}
\usepackage{booktabs}
\usepackage[multiple]{footmisc}

\tikzstyle{arrow} = [thick,->,>=stealth]

\tikzstyle{rounded_box} = [rectangle, rounded corners,
minimum width=2cm, 
minimum height=1cm, 
text centered, 
text width=2cm, 
draw=black,
fill=red!30]

\tikzstyle{box} = [rectangle, 
minimum width=3.5cm, 
minimum height=1cm, 
text centered, 
text width=3.5cm, 
draw=black,
fill=blue!30]

\tikzstyle{result} = [ellipse, 
minimum width=1.5cm, 
minimum height=1cm, 
text centered, 
text width=1.5cm, 
draw=black,
fill=green!30]

\usepackage{csquotes}

\begin{document}

\title{Analysis of real-space transport channels for electrons and holes in halide perovskites}

\author{Frederik Vonhoff}
\affiliation{%
 Physics Department, TUM School of Natural Sciences, Technical University of Munich, 85748 Garching, Germany
 }
\author{Maximilian J. Schilcher}
\affiliation{%
 Physics Department, TUM School of Natural Sciences, Technical University of Munich, 85748 Garching, Germany
 }
\author{David R. Reichman}
\affiliation{%
Department of Chemistry, Columbia University, New York, New York 10027, USA
}
\author{David A. Egger}
\email{david.egger@tum.de}
\affiliation{%
 Physics Department, TUM School of Natural Sciences, Technical University of Munich, 85748 Garching, Germany
}%
\affiliation{%
 Atomistic Modeling Center, Munich Data Science Institute, Technical University of Munich, 85748 Garching, Germany
 }
\date{\today}
             
\begin{abstract}
Predicting and explaining charge carrier transport in halide perovskites is a formidable challenge because of the unusual vibrational and electron-phonon coupling properties of these materials.
This study explores charge carrier transport in two prototypical halide perovskite materials, \ce{MAPbBr3} and \ce{MAPbI3}, using a dynamic disorder model. 
Focusing on the role of real-space transport channels, we analyze temporal orbital occupations to assess the impact of material-specific on-site energy levels and spin-orbit coupling (SOC) strengths. 
Our findings reveal that both on-site energies and SOC magnitude significantly influence the orbital occupation dynamics, thereby affecting charge dispersal and carrier mobility. 
In particular, energy gaps across on-site levels and the halide SOC strength govern the filling of transport channels over time. 
This leads us to identify the $pp\pi$ channel as a critical bottleneck for charge transport and to provide insights into the differences between electron and hole transport across the two materials.
\end{abstract}


\maketitle


\section{\label{sec:level0}Introduction}

Halide perovskites (HaPs) are an interesting class of semiconducting materials for optoelectronic applications, such as solar cells and light-emitting devices \cite{review_devices,review_solarcell,review_opto_devices}.
The exceptional performance of HaPs in device applications stems from their advantageous optoelectronic properties \cite{transport_properties,intrguing_optoelectronic,HaP_performance,review_HaP_properties}.
Prototypical HaPs exhibit direct band gaps, almost ideal absorption features with sharp absorption edges,\cite{absorption} relatively low exciton binding energies, \cite{exciton_binding}, and long charge carrier diffusion lengths \cite{diffusion_length2,diffusion_length}.
Furthermore, their \ce{ABX3} stoichiometry allows for realizing multiple material compositions with tunable optoelectronic properties.
For example, the fundamental band gap of HaPs can be modified via their ionic composition,\cite{bandgap_tuning,intrguing_optoelectronic,HaP_performance}, which is critical for utilizing these materials in tandem solar cells.

The design of new HaP materials hinges on improving our microscopic understanding of charge carrier transport in these systems, as it is a foundational mechanism in all semiconductor devices.
However, modeling transport in HaPs defies established textbook frameworks at the same time:
carrier mobilities exhibit power-law behavior, which are key signatures of band-like transport reminiscent of classical inorganic semiconductors such as bulk Si and GaAs \cite{Cardona_Yu}.
Simultaneously, however, carrier relaxation times in HaPs are ultrafast and mean-free paths ultrashort,\cite{MaxPaper1} which is reminiscent of  molecular semiconductors \cite{Troisi2006,Fratini2009,beyond_polaron,Blumberger2020}.
Related to this, these materials show strongly anharmonic vibrations that cause large fluctuations in their electronic properties and unusual optoelectronic behavior \cite{Quarti2016,Yaffe2017,Marronnier2018,Mayers,Gehrmann2019,MaxPaper2,Seidl2023,Zacharias2023,Frenzel2023,Park2023,Wiktor2023,HyltonFarrington2024,Zhu2025}.
Predictive computational modeling of carrier transport of such seemingly contradicting scenarios is challenging, especially because violations of the Mott-Ioffe-Regel criterion \cite{Ioffe_Regel,Mott} prohibit using standard band-transport theories \cite{Fratini2009,beyond_polaron,MaxPaper1}.
Specifically, when mean-free paths of charge carriers approach or even drop below the wavelengths of their wave packets, a perturbative treatment of carrier-lattice coupling is not sensible as band-like quasiparticles are not well-defined anymore.
Indeed, established frameworks employing simplified models of Fröhlich polarons in combination with the Boltzmann transport equation fail to predict experimentally-measured transport characteristics\cite{Mayers}, which is expected for the parameter regime of HaPs\cite{beyond_polaron}.
Development of new predictive models for charge transport in HaPs is therefore an important challenge for theory.
Such models can generate a microscopic understanding of carrier transport in HaPs and faster progress and development of design principles in the context of other materials families. 
These include inorganic semiconductors such as oxides \cite{Wiktor2017,SrTiO3} and nitrides \cite{FranziskaPaper} as well as organic semiconductors \cite{Troisi2006,Fratini2009,Beljonne2011,Troisi_2011,Fratini2012,fratini_transient_2016,oberhofer_charge_2017,Blumberger2019,Blumberger2020,Egger_Yaffe_2020,Beljonne_Blumberger_2023}.

To tackle the illustrated challenges of modeling carrier transport in HaPs, some of the authors proposed a dynamic disorder (DD) model \cite{Mayers,MaxPaper1,MaxPaper2}.
This model incorporates vibrational anharmonicity via the use of molecular dynamics (MD), which is critical to capturing the pronounced vibrational anharmonicity. 
Transient localization of charge carriers accompanying the vibrational fluctuations \cite{MaxPaper1,fratini_transient_2016,Lacroix2020} is modeled via a tight binding (TB) Hamiltonian that explicitly depends on the atomic configurations obtained in MD.
The TB model is parameterized from first principles with density functional theory (DFT) using the geometries from MD snapshots.
The resulting time-dependent Hamiltonian allows one to predict the time evolution of electron (e$^-$) and hole (h$^+$) wave functions in the DD model.
The computed carrier mobilities have been found to be in broad agreement with experiments regarding magnitudes, temperature dependencies, and chemical trends \cite{Mayers,MaxPaper2}.
Notably, a recent experimental study provided evidence in support of the DD model, highlighting the crucial role of DD as the dominant scattering mechanism in HaPs \cite{Dörflinger2025}.

In this study, we further develop the DD model to enable real-space analysis of carrier dynamics and investigate the consequences of using more accurate electronic-structure calculations to parameterize it.
Using the \ce{MAPbBr3} and \ce{MAPbI3} variants, we focus on how on-site energy levels and the magnitude of spin-orbit coupling (SOC), both determined by the chemical composition of the materials, conspire with the dynamic fluctuations of the electronic structure in HaPs to produce the novel transport characteristics of these materials. 
To this end, we resolve time-dependent orbital occupations and elucidate the carrier-scattering mechanisms in three real-space transport channels that together determine carrier transport by either enhancing or suppressing mobility.
By providing an in-depth analysis of transport channels in HaPs, this work enables improved microscopic understandings of the foundational properties of these semiconducting materials.

\section{\label{sec:level1}Methods}

\begin{figure}[b]
\begin{tikzpicture}[node distance=2.7cm]

\node (A) [box] {Nuclear dynamics and electronic structure from first principles};
\node (aiMD) [rounded_box, left of=A, xshift=-2.3cm] {Ab initio molecular dynamics};
\node (B) [box, below of=A] {Large scale structural and electronic dynamics};
\node (FFMD) [rounded_box, left of=B, xshift=-2.3cm] {Force field molecular dynamics};
\node (C) [box, below of=B] {Time evolution of carrier wave functions};
\node (mu) [result, left of=C, xshift=-2.3cm] {Carrier dynamics};

\draw [arrow] (aiMD) -- node[anchor=south] {DFT \&} node[anchor=north] {\texttt{Wannier90}}  (A);
\draw [arrow] (FFMD) -- node[anchor=south] {nuclear} node[anchor=north] {trajectories}  (B);
\draw [arrow] (A) -- node[anchor=east] {Mapping rules} node[anchor=west] {for TB model \phantom{phantom}} (B);
\draw [arrow] (B) -- node[anchor=east] {Time series of} node[anchor=west] {Hamiltonians} (C);
\draw [arrow] (C) -- node[anchor=south] {orbital} node[anchor=north] {population} (mu);

\end{tikzpicture}
\caption{Workflow of the dynamic disorder (DD) model: each blue rectangular box represents one conceptual step in parameterizing the model and performing calculations with it, see text for details.}
\label{fig:workflow}
\end{figure}
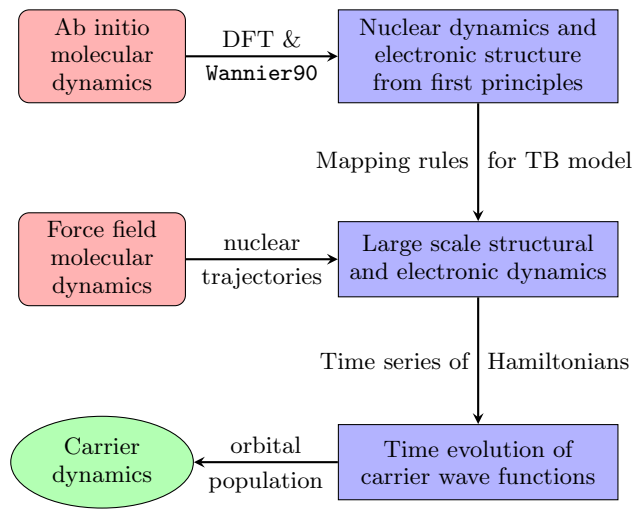

Fig.~\ref{fig:workflow} illustrates the workflow of the DD model developed in refs.~\cite{Mayers,MaxPaper2}.
In brief, it involves three key steps: 
First, we sample nuclear dynamics and electronic structure with ab initio MD to parameterize TB Hamiltonians.
Second, TB Hamiltonians for large scale structural and electronic dynamics are constructed.
And third, carrier wave packets are evolved in time to monitor scattering effects due to nuclear dynamics and extract transport characteristics from it.
We describe these three main steps in the following.

\subsection{\label{sec:level1a}Nuclear dynamics and electronic structure from first principles}

To parameterize the TB Hamiltonians, we create samples of the nuclear dynamics and electronic structures at finite temperature using ab initio MD.
Specifically, we use data from ref.~\cite{MaxPaper2}, comprising MD trajectories spanning \SI{35}{\pico\second} for \ce{MAPbI3} and \ce{MAPbBr3} at \SI{300}{\kelvin} in $2\times2\times2$ supercells. 
Extracting roughly 100 snapshots, we perform DFT-based electronic-structure calculation using the Vienna Ab initio Simulation Package (\texttt{VASP}) \cite{VASP} with the hybrid Heyd–Scuseria–Ernzerhof (HSE) functional \cite{HSE,Erratum_HSE} incorporating SOC.
Previous studies with the DD model \cite{Mayers,MaxPaper2} used the Perdew–Burke–Ernzerhof (PBE) functional \cite{PBE}, which captures the main features in electronic structures of HaPs but is less accurate than HSE \cite{bandgap_2015,bandgap_PBE_HSE}.

The TB parameters for on-site energies, hoppings, and SOC are extracted via \texttt{Wannier90} \cite{W90_2} for Pb and X ions, using Pb-$s$ and $p$, as well as halide X-$p$ atomic orbitals as basis.
We denote on-site energies as $\varepsilon^i_\alpha$, where $i$ is the atomic index and $\alpha$ the orbital index, and SOC parameters for X and Pb ions as $\gamma_\mathrm{X}$ and $\gamma_\mathrm{Pb}$, respectively.
We include the three hopping elements $t_{sp\sigma}$, $t_{pp\sigma}$, and $t_{pp\pi}$, which are indexed according to bonding type.

When hopping elements are plotted against interatomic Pb-X distances, we find that the HSE data broadly confirm previous PBE-based results \footnote{See Supplemental Material at LINK for more details on the impact of the functional on the hopping elements and the connection of the Crystal Orbital Bond Index to the hopping elements.}.
The data show the expected exponential decrease with increasing distance, which can be fit by a function of the form $a\cdot\exp(-b\cdot x)+c$. 
This establishes a mapping rule between results from DFT calculations and TB parameters.
We find that the fitting parameters $b$ and $c$ are similar for PBE and HSE calculations. 
However, the parameter $a$ is significantly larger in case of HSE compared to PBE data, which results in larger values for the hopping elements.
For the on-site energies, the mapping is achieved via Ewald summations, where the computed parameters are shifted such that their mean values align with those calculated through DFT and \texttt{Wannier90} for \ce{MAPbI3} and \ce{MAPbBr3}, respectively.
The SOC parameters for Pb and X ions are set to their respective mean values that we calculate with \texttt{Wannier90} across the MD trajectory.
In summary, the extracted hopping elements, on-site energies, and SOC parameters are combined into a first-principles database, which provides a mapping rule to construct Hamiltonian matrices of large supercells depending on the time-dependent material geometries we extract from MD.

\subsection{\label{sec:level1b}Large scale structural and electronic dynamics}

We construct Hamiltonian matrices for large supercells with the TB mapping rule parameterized from first principles.
To model the structural dynamics at different temperatures, we use force field MD data from ref.~\cite{MaxPaper2}.
They contain $16\times16\times16$ supercells of \ce{MAPbI3} and \ce{MAPbBr3} at five different temperatures spanning \SIrange{200}{350}{\kelvin} from which we extract 100 snapshots corresponding to \SI{100}{\femto\second}.
Using these snapshots, we obtain time series of Hamiltonians with the mapping rule using atomic distances as input and assigning on-site terms as well as SOC parameters as described before.
To prevent finite-size effects, we periodically continue Hamiltonians of $16\times16\times16$ cells into $96\times96\times96$ cells.

\subsection{\label{sec:level1c}Time evolution of carrier wave functions}

We compute the time evolution of differently initialized wave functions from the large scale structural and electronic dynamics. 
For initialization of e$^-$ and h$^+$ carriers, we calculate the electronic Bloch eigenstates from the Hamiltonian of the first snapshot.
These states are constructed from the primitive unit cell calculated on an $8\times8\times8$ $k$-grid, and utilize Bloch eigenstates corresponding to the conduction band for e$^-$ and the valence band for h$^+$, respectively. 
Transforming them to maximally localized Wannier functions within a $8\times8\times8$ supercell is achieved via \texttt{Wannier90}.
We neglect orbitals that do not strongly contribute to e$^-$ and h$^+$ wave functions in this procedure, that is, we employ Pb-$p$ orbitals for e$^-$ and Pb-$s$ as well as X-$p$ orbitals for h$^+$.

The Wannier functions for e$^-$ or h$^+$ are then placed at different starting positions in the original $16\times16\times16$ supercell such that the initial wave function always starts evolving in the center of the periodically continued $96\times96\times96$ supercell.
We propagate these initial wave functions via a product of time evolution operators, where each operator is constructed out of one Hamiltonian from the temporal series.
From the time-evolved wave functions, we calculate the mean squared displacement (MSD) of charge carriers.
Its diffusive part, averaged over different starting positions of the initial wave function in the crystal, yields e$^-$ and h$^+$ mobilities via the Einstein-Smoluchowski relation.
From this procedure, one can determine the carrier mobility at different temperatures \cite{Mayers,MaxPaper2}.

\section{\label{sec:level2}Results}

In this study, we investigate the interplay between on-site energies, the magnitude of SOC, and carrier dynamics in \ce{MAPbBr3} and \ce{MAPbI3}.
To this end, we compute dynamic orbital populations and assess their role in electronic transport phenomena via a real-space analysis of carrier scattering.

\subsection{\label{sec:level2a}Initial wave functions localized at single orbitals}

\begin{figure}
\includegraphics{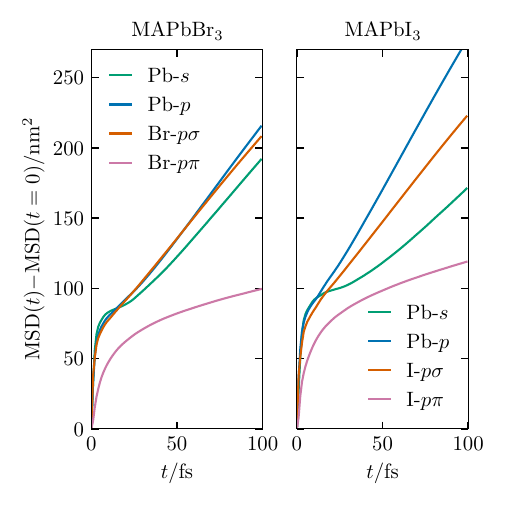} 
\caption{Mean-squared displacement of carrier wave functions initialized in either a Pb-$s$, Pb-$p$, X-$p\sigma$, or X-$p\pi$ orbital for \ce{MAPbBr3} and \ce{MAPbI3}.}
\label{fig:avgMSD_pureOrb}
\end{figure}

We start our investigation by calculating the dynamic evolution of wave functions localized at an individual orbital that contributes to band-edge states of \ce{MAPbI3} and \ce{MAPbBr3}.
These are the Pb-$s$, Pb-$p$, X-$p\sigma$, and X-$p\pi$ orbitals.
Here, X-$p\pi$ refers to orbitals involved in $pp\pi$-hopping elements due to interactions between X-$p$ and Pb-$p$ states.
Likewise, X-$p\sigma$ corresponds to orbitals involved in $sp\sigma$- and $pp\sigma$-hopping elements.
Fig.~\ref{fig:avgMSD_pureOrb} shows their MSDs resulting from an initial wave function localized at a single orbital on one atomic site for the two HaPs.
Except for the X-$p\pi$ orbitals, the MSDs can be separated into three regimes.
First, the MSD rises sharply during {$\approx$}\SI{10}{\femto\second}, reflecting a quick dispersal of the artificially localized initial wave function.
Second, between roughly \SIrange{10}{35}{\femto\second}, the MSDs show a quadratic increase in time, i.e., wave functions spread ballistically.
Third, for times larger than {$\approx$}\SI{35}{\femto\second}, we find a linear increase in the MSDs and diffusive transport sets in.

In contrast, the dispersal of X-$p\pi$ orbitals does not follow this sequence and is strongly suppressed.
We observe that MSDs for these orbitals saturate at a comparably low value in both materials, indicating that an entirely different transport mechanism is at play in this case.
Comparing the transport behavior for the other orbital types, in case of \ce{MAPbBr3} the MSDs are close to each other while for \ce{MAPbI3} they are more spread with an enhanced Pb-$p$ and a suppressed Pb-$s$ curve.
We conclude that the orbital configuration influences the transport mechanisms within the DD model resulting in different spread of the carrier wave function at early times of the system's quantum dynamics.

\begin{figure}
\includegraphics{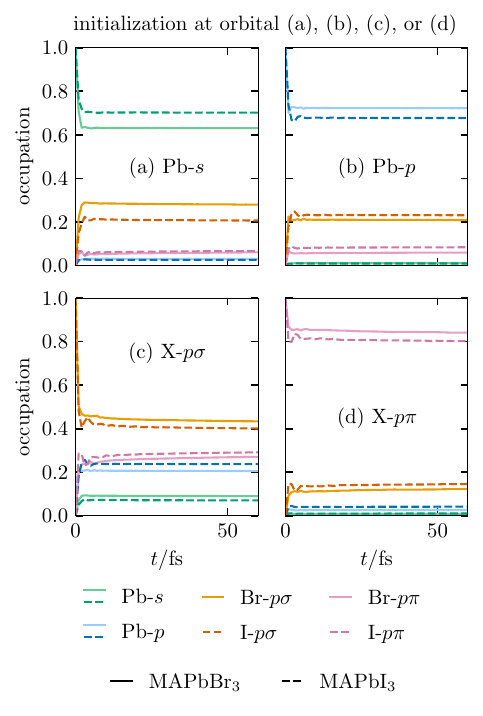} 
\caption{Time evolution of relative occupation for the orbital types Pb-$s$, Pb-$p$, X-$p\sigma$ and X-$p\pi$ in 
\ce{MAPbX3} (X=Br, I) when different wave functions localized in a single orbital are used in the initialization (cf. Fig.~\ref{fig:avgMSD_pureOrb}). Subplots (a) - (d) correspond to an initial wave function localized in a single Pb-$s$ (a), Pb-$p$ (b), X-$p\sigma$ (c), and X-$p\pi$ orbital (d).}
\label{fig:occupation_pureOrb}
\end{figure}

Fig.~\ref{fig:occupation_pureOrb} shows the population across different orbitals as a function of time proceeding the initialization of wave functions we showed in Fig.~\ref{fig:avgMSD_pureOrb}.
We observe sharp changes among relative orbital occupations during the first few femtoseconds for all initialization scenarios. 
They correspond to the strong dispersal of the artificially localized initial wave functions seen at early times in the MSDs of Fig.~\ref{fig:avgMSD_pureOrb}.
This behavior is followed by plateaus of relative orbital occupation in time when local fluctuations are averaged out.
These stable orbital occupations reflect a balanced transport mechanism within the DD model, in which each orbital type steadily contributes to overall charge conduction.

Like above, we observe differences depending on which orbital is used in the initialization of wave functions:
when a \mbox{Pb-$s$} state is chosen in the initialization (see Fig.~\ref{fig:occupation_pureOrb}a), Pb-$s$ and X-$p\sigma$ orbitals dominate the temporal charge carrier distribution. 
Notably, these orbital types are involved in $sp\sigma$ bonds. 
In contrast, when a Pb-$p$ state is used in the initialization (see Fig.~\ref{fig:occupation_pureOrb}b), charge carriers are primarily occupying Pb-$p$ and X-$p\sigma$ orbitals over time. 
These orbital types are involved in $pp\sigma$ bonds.

\begin{figure}
\includegraphics[width=0.5\textwidth]{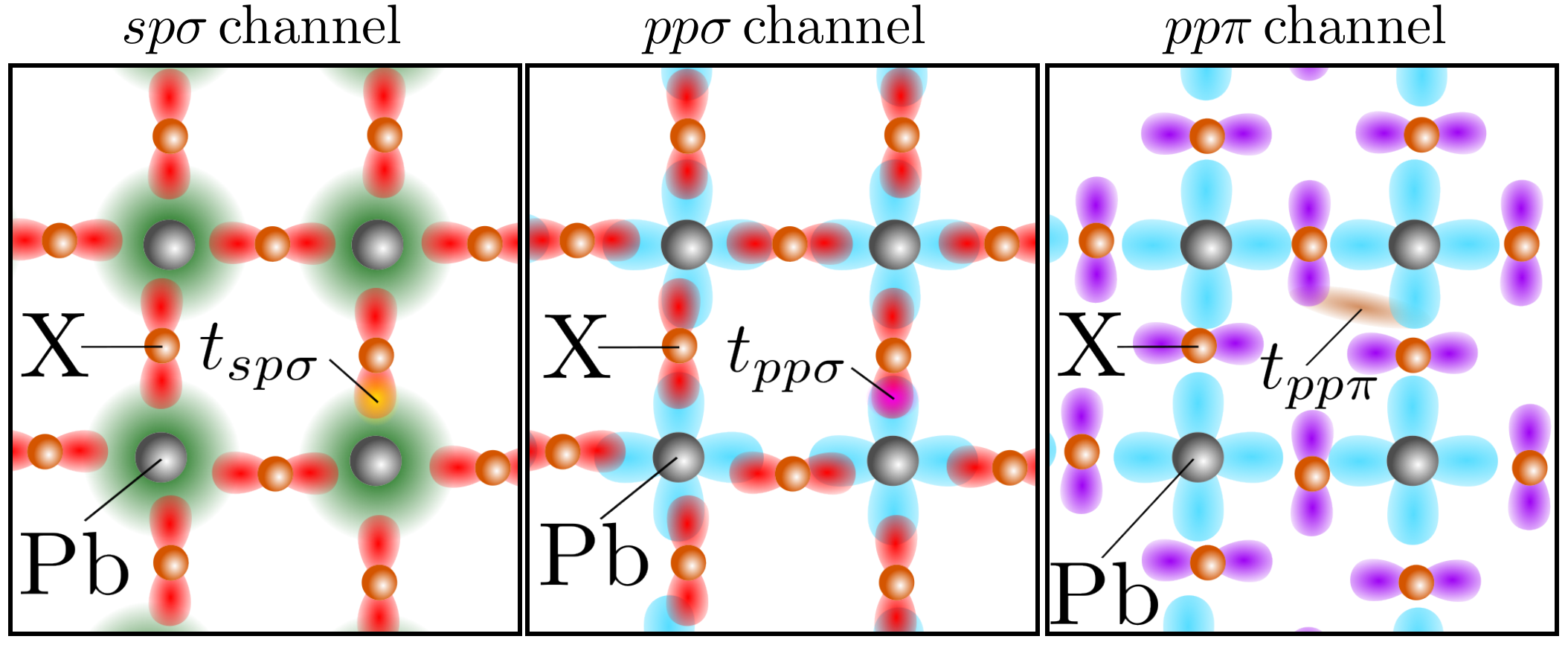} 
\caption{Schematic visualization of the three transport channels, each driven by one of the TB hopping parameter types $sp\sigma$, $pp\sigma$, and $pp\pi$. Pb ions are shown in gray and halides in orange, Pb-$s$ orbitals in green, Pb-$p$ orbitals in blue, X-$p\sigma$ orbitals in red, and X-$p\pi$ orbitals in purple. For easy visualization, we show a two-dimensional slice of a cubic HaP structure without MA cations.}
\label{fig:transport_channels}
\end{figure}

We rationalize the above findings using real-space transport channels that are visualized schematically in Fig.~\ref{fig:transport_channels}.
Each channel corresponds to one of the three bond types, i.e., $sp\sigma$, $pp\sigma$, and $pp\pi$. 
In the DD model, charge carrier wave functions are transferred exclusively through the orbitals involved in these bonds and, hence, the channels collectively determine the charge carrier transport in HaPs. 
Revisiting the above results, in Fig.~\ref{fig:occupation_pureOrb}a only the $sp\sigma$ channel (Fig.~\ref{fig:transport_channels}, left panel) is active because only Pb-$s$ and X-$p\sigma$ orbitals are significantly occupied.
Likewise, in Fig.~\ref{fig:occupation_pureOrb}b, the $pp\sigma$ channel (Fig.~\ref{fig:transport_channels}, middle panel) almost exclusively determines transport as mainly Pb-$p$ and X-$p\sigma$ orbitals are occupied. 
However, a mixture of all channels occurs in Fig.~\ref{fig:occupation_pureOrb}c because the initial wave function localized at an X-$p\sigma$ orbital connects both the $sp\sigma$ and $pp\sigma$ channels.

In contrast, Fig.~\ref{fig:occupation_pureOrb}d shows that carriers remain localized in X-$p\pi$ orbitals. 
We compute averages and standard deviations of all TB parameters along an MD trajectory of \ce{MAPbBr3} and \ce{MAPbI3} at \SI{300}{\kelvin} and report them in Tab.~\ref{tab:TB_params}. 
Compared to $\bar{t}_{sp\sigma}$ and $\bar{t}_{pp\sigma}$, the hopping parameter $\bar{t}_{pp\pi}$ is significantly smaller.
This can be understood by considering the reduced overlap of orbitals involved in the $pp\pi$ channel (Fig.~\ref{fig:transport_channels}, right panel). 
Consequently, the $pp\pi$ channel is a bottleneck channel for charge transport, which explains the reduced MSD we found when X-$p\pi$ orbitals were used to initialize the wave function (cf. Fig.~\ref{fig:avgMSD_pureOrb}).

\begin{table}[h!]
\centering
\caption{Mean values of on-site, hopping, and SOC parameters for \ce{MAPbBr3} and \ce{MAPbI3} at \SI{300}{\kelvin}. The values are calculated by averaging all parameters of the same type within the Hamiltonian time series. We also report standard deviations for all parameters except the ones for SOC, which we have set to their mean values (see methods section).}
\vspace{2mm}
\begin{tabularx}{0.48\textwidth}{c|>{\centering\arraybackslash}X|>{\centering\arraybackslash}X}
    \toprule
        {} & \ce{MAPbBr3} & \ce{MAPbI3} \\
    \midrule
        $\bar{\varepsilon}_\mathrm{Pb}^p$ & \SI[separate-uncertainty = true]{3.23(20)}{\electronvolt} & \SI[separate-uncertainty = true]{2.58(20)}{\electronvolt} \\
                                    &       &      \\ 
        $\bar{\varepsilon}_\mathrm{X}^p$ & \SI[separate-uncertainty = true]{-2.06(25)}{\electronvolt} & \SI[separate-uncertainty = true]{-1.57(23)}{\electronvolt} \\ 
                                    &       &      \\ 
        $\bar{\varepsilon}_\mathrm{Pb}^s$ & \SI[separate-uncertainty = true]{-6.74(20)}{\electronvolt} & \SI[separate-uncertainty = true]{-7.06(19)}{\electronvolt} \\  
    \midrule
        $\bar{t}_{sp\sigma}$ & \SI[separate-uncertainty = true]{1.25(28)}{\electronvolt} & \SI[separate-uncertainty = true]{1.12(25)}{\electronvolt} \\
                       &       &      \\ 
        $\bar{t}_{pp\sigma}$ & \SI[separate-uncertainty = true]{2.01(29)}{\electronvolt} & \SI[separate-uncertainty = true]{1.90(26)}{\electronvolt} \\
                       &       &      \\ 
        $\bar{t}_{pp\pi}$ & \SI[separate-uncertainty = true]{0.48(13)}{\electronvolt} & \SI[separate-uncertainty = true]{0.42(12)}{\electronvolt}  \\ 
    \midrule
        $\bar{\gamma}_\mathrm{X}$ & \SI[separate-uncertainty = true]{0.165}{\electronvolt} & \SI[separate-uncertainty = true]{0.323}{\electronvolt} \\
        &       &      \\ 
        $\bar{\gamma}_\mathrm{Pb}$ & \SI[separate-uncertainty = true]{0.538}{\electronvolt} & \SI[separate-uncertainty = true]{0.520}{\electronvolt} \\
    \bottomrule
\end{tabularx}
\label{tab:TB_params}
\end{table}

\subsection{\label{sec:level2b}Full initial electron and hole wave functions}

In this section, we analyze the temporal evolution of MSDs and relative orbital occupations using full e$^-$ and h$^+$ wave functions and apply the insights about real-space transport channels from above.
Furthermore, we study how on-site levels and SOC affect the transport channels to explain differences between \ce{MAPbBr3} and \ce{MAPbI3}. 
We further demonstrate a tuning of charge transport by adjusting on-site levels and SOC parameters.

\begin{figure}
\includegraphics{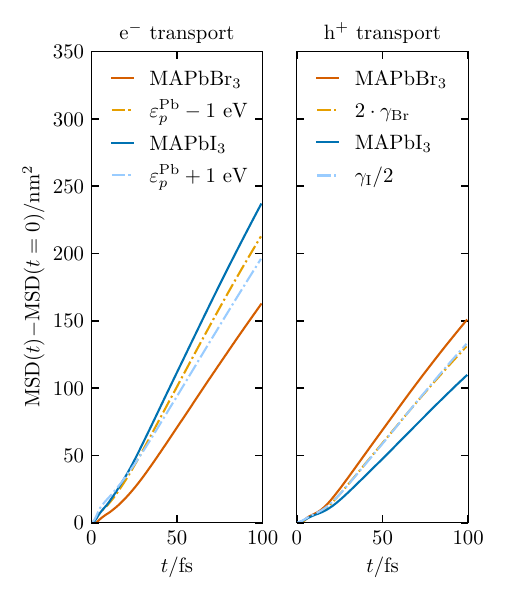} 
\caption{MSDs for full initialization of electron e$^-$ (left) and hole h$^+$ wave functions (right). Solid lines show results of calculations performed as described in the methods sections without altering TB parameters. Dashed lines show results of calculations in which the TB parameters have been modified. 
}
\label{fig:avgMSD}
\end{figure}

Fig.~\ref{fig:avgMSD} shows the MSD over time for full initial e$^-$ and h$^+$ wave functions (see methods section).
Like above, the MSDs display three temporal regimes, beginning with a sharp rise, followed by a ballistic quadratic increase, and merging into a diffusive linear behavior.
Compared to Fig.~\ref{fig:avgMSD_pureOrb}, the initial sharp rise is significantly less pronounced, owing to the more physical choice of initial wave functions.

We explore the role of on-site energies on charge transport by modifying the TB parameters, shifting the on-site energy levels of Pb-$p$ orbitals, $\varepsilon_p^\mathrm{Pb}$, for e$^-$ transport (Fig.~\ref{fig:avgMSD}, left panel). 
We separately rescale the halide SOC strengths, $\gamma_\mathrm{X}$, for h$^+$ transport (Fig.~\ref{fig:avgMSD}, right panel).
Without these parameter changes, e$^-$ transport is stronger than h$^+$ transport in both materials.
Moreover, for e$^-$ transport the MSD increases more strongly in \ce{MAPbI3} compared to \ce{MAPbBr3}, and vice versa for h$^+$ transport.
Exploring the origins of these observations, we find that modifying the Pb-$p$ on-site energy levels enhances the magnitude of the MSD for e$^-$ when $\varepsilon_p^\mathrm{Pb}$ is lowered, while raising $\varepsilon_p^\mathrm{Pb}$ has the opposite effect.
We observe complementary effects for adjusting the halide onsite levels, $\varepsilon_p^\mathrm{X}$, such that a smaller gap between the two levels, $\varepsilon_p^\mathrm{Pb}$ and $\varepsilon_p^\mathrm{X}$, always intensifies the overall transport, while a larger gap weakens it.
In contrast, altering the on-site energy levels of Pb-$s$ orbitals, $\varepsilon_s^\mathrm{Pb}$, barely influences the e$^-$ transport.
However, for h$^+$ transport, reducing and increasing $\varepsilon_p^\mathrm{Pb}$ raise the MSD magnitude by similar amounts.
Meanwhile, an increased/decreased halide SOC strength, $\gamma_\mathrm{X}$, suppresses/enhances the magnitude of the MSD for h$^+$ transport, whereas rescaling $\gamma_\mathrm{X}$ has almost no impact on e$^-$ transport.

\begin{figure}
\includegraphics{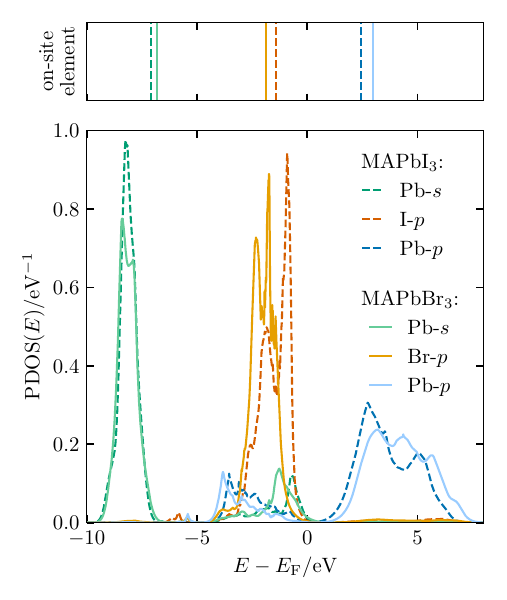} 
\caption{Projected density of states (PDOS) for the orbital types Pb-$s$, Pb-$p$, X-$p$ in \ce{MAPbBr3} (solid lines) and \ce{MAPbI3} (dashed lines). The eigenenergies were evaluated with DFT using a $2\times2\times2$ supercell and the HSE functional. 
We used \texttt{Wannier90} to calculate the PDOS and on-site level energies, and set the Fermi energy, $E_F$, to the energy of the highest occupied state.}
\label{fig:DOS}
\end{figure}

To rationalize the consequences of the parameter changes on the transport results, we link the tight binding model to the more intuitive projected density of states (PDOS) and illustrate the impact of level shifts on the electronic structure.
Therefore, we use DFT to calculate the PDOS of \ce{MAPbBr3} and \ce{MAPbI3} for the three orbitals types, Pb-$s$, Pb-$p$, and X-$p$, at one representative atomic site along the MD trajectory (Fig.~\ref{fig:DOS}, lower panel) and calculate their corresponding on-site energy levels (Fig.~\ref{fig:DOS}, upper panel).
As is well established for these materials \cite{Ashhab2017,Mehdizadeh2019}, Pb-$p$ orbitals dominate the conduction band while the valence band shows contributions of all orbital types with X-$p$ orbitals being most prominent.
It is also well-known that the fundamental band gap of \ce{MAPbI3} is smaller than that of \ce{MAPbBr3} \cite{bandgap_PBE_HSE}.
This is also reflected in their on-site level structure since on-site levels match the main weights of their corresponding PDOS.
Moreover, the PDOS relates analytically to the energy of the on-site energy levels via 
\begin{equation}
    \varepsilon_\alpha^{i} = \int_{-\infty}^\infty \varepsilon \cdot \mathrm{PDOS}_\alpha^{i}(\varepsilon) \: \mathrm{d}\varepsilon.
    \label{eqn:onsite_PDOS}
\end{equation}
This relation directly connects $\mathrm{PDOS}_\alpha^{i}(\varepsilon)$, that is the PDOS of an orbital $\alpha$ at site $i$, to the respective on-site energy, $\varepsilon_\alpha^{i}$.
Likewise, the crystal orbital bond index (COBI) is related to hopping elements \footnotemark[\value{footnote}].
Together, the PDOS and the COBI offer an easy-to-follow framework for interpreting the impact of changes in the tight binding parametrization on the electronic structure.

\begin{figure}
\includegraphics{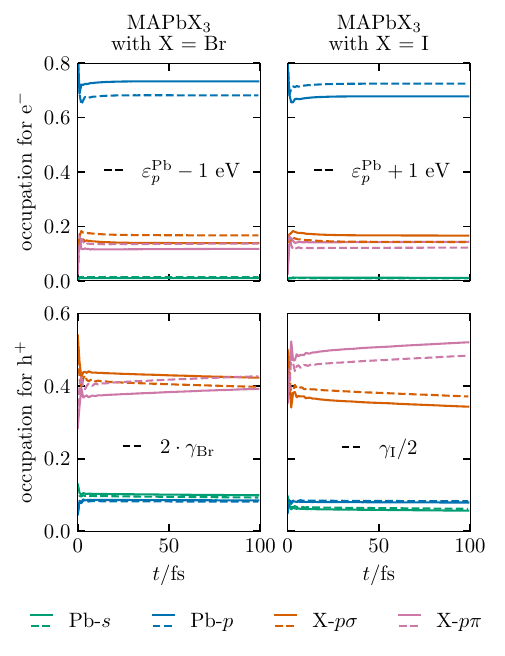} 
\caption{
Time evolution of relative occupation for the orbital types Pb-$s$, Pb-$p$, X-$p\sigma$ and X-$p\pi$ in 
\ce{MAPbX3} (X=Br, I) when full initial electron e$^-$ (top panels) and hole h$^+$ wave functions (bottom panels) are used in the initialization (cf. Fig.~\ref{fig:avgMSD}). 
Each subplot reflects the parameter change specified in its legend.}
\label{fig:occupation}
\end{figure}

To further investigate the parameter adjustments shown in Fig.~\ref{fig:avgMSD}, we analyze the relative orbital occupation over time in Fig.~\ref{fig:occupation} for e$^-$ and h$^+$ wave functions that are fully initialized according to the Bloch eigenstates of conduction- and valence-band states.
We probe the role of the transport channels described above in realistic carrier-scattering scenarios and determine how shifts of on-site levels and halide SOC magnitude impact transport.

The upper two panels of Fig.~\ref{fig:occupation} show relative orbital occupations for initial e$^-$ wave functions with and without introducing shifts for $\varepsilon_p^\mathrm{Pb}$ (cf. Fig.~\ref{fig:avgMSD}, left panel).
In both HaPs, initial e$^-$ wave functions are mostly composed of orbitals that make up the conduction band states, i.e., they show dominant Pb-$p$ contributions (cf. Fig.~\ref{fig:DOS}). 
These orbitals remain the primary contributors over time, whereas the Pb-$s$ orbitals remain weakly occupied.
Since the Pb-$s$ orbital is essential in the composition of the $sp\sigma$ channel, the lack of its occupation indicates that the $sp\sigma$ channel is irrelevant in case of e$^-$ transport.
Thus, on-site level shifts for Pb-$s$ orbitals do not affect e$^-$ transport as described above. 
However, level shifts for Pb-$p$ orbitals have a noticeable effect: when we apply shifts that align the Pb-$p$ on-site levels for the two HaPs more (cf. Tab.~\ref{tab:TB_params}), we find the relative occupations between \ce{MAPbBr3} and \ce{MAPbI3} to become more similar too.
This is also reflected in the occupation of Pb-$p$ levels being closer to X-$p\sigma$ and X-$p\pi$ orbitals when we reduce the energy level of Pb-$p$, as shown in the upper left panel of Fig.~\ref{fig:occupation}.

The on-site level shifts and related changes in the occupations influence the transport strength through the following mechanism.
When we shift two on-site energy levels involved in a transport channel towards each other, their occupations also become slightly more aligned.
This more balanced charge distribution among the two involved orbital types implies a more steady flow of the carrier wave function and, thereby, leads to enhanced carrier transport along this channel.
In essence, a smaller energy gap between the Pb-$p$ and halide on-site levels facilitates level transitions and, therefore, stronger transport along the $pp\sigma$ channel, which is consistent with our observations (cf. Fig.~\ref{fig:avgMSD}, left panel).
The stimulation of transitions by level shifts found here is reminiscent of the proportionality between transition probability and energy level difference for Rabi cycles, which can be analytically shown for a two-level system \cite{Atkins_Friedman}.
The same principles apply when we introduce on-site level shifts for X-$p$ orbitals.
Accordingly, the differences in X-$p$ and Pb-$p$ on-site level gaps between the two HaPs partly account for the higher MSD for \ce{MAPbI3} compared to \ce{MAPbBr3} in case of e$^-$ transport.
The residual difference in their MSDs, even after aligning the on-site level structure, then predominantly stems from the variation in the strengths and DD of the hopping parameters (cf. Tab.~\ref{tab:TB_params}).
We note in passing that the interplay between level gaps we describe here is different from the impact of varying the Holstein electron-phonon coupling, even though it can also manifest in shifts of the on-site levels \cite{Berkelbach2020}, and also different from the influence of the DD originating from the on-site elements \cite{Lacroix2020}.
We further note that \ce{MASnI3} shows a narrower on-site level gap between I-$p$ and metal cation $p$-orbitals \cite{Paschal2021} compared to \ce{MAPbI3}, which could at least partly explain the generally larger carrier mobilities of \ce{MASnI3} compared to \ce{MAPbI3} \cite{Herz2017}.

The lower two panels of Fig.~\ref{fig:occupation} show relative orbital occupations for initial h$^+$ wave functions with and without rescaling of $\gamma_\mathrm{X}$ (cf. Fig.~\ref{fig:avgMSD}, right panel).
In both HaPs, X-$p$ states are the strongest contributing orbital types owing to their large weights in the valence band states (cf. Fig.~\ref{fig:DOS}).
Since X-$p$ orbitals participate in every channel, a mixture of all three channels occurs in case of h$^+$ transport.
Hence, the bottleneck, $pp\pi$ channel also plays an active role in case of h$^+$ transport and weakens it.
This can be assessed comparing the magnitude of the X-$p\pi$ orbital occupation for h$^+$ and e$^-$, which in case of the latter is much less pronounced and, thus, the suppressing nature of the $pp\pi$ channel is substantially less relevant.

Comparing the two HaPs for the case of h$^+$ transport, a stronger occupation of X-$p\pi$ orbitals for \ce{MAPbI3} than for \ce{MAPbBr3} entails a more dominant role of the $pp\pi$ channel, which impedes transport.
Fig.~\ref{fig:occupation} shows that the occupation of X-$p\pi$ orbitals is impacted by the magnitude of $\gamma_\mathrm{X}$, which can be interpreted as an additional hopping term between X-$p\pi$ and X-$p\sigma$ orbitals within our tight binding model.
As indicated by the two lower panels of Fig.~\ref{fig:occupation}, increasing $\gamma_\mathrm{X}$ increases the occupation of X-$p\pi$ orbitals and suppresses transport.
For the same reason, e$^-$ transport is unaffected by $\gamma_\mathrm{X}$ because the occupation of X-$p\pi$ is negligible in this case.
Since halide SOC parameters of \ce{MAPbI3} are approximately twice as large as that of \ce{MAPbBr3} (see Tab.~\ref{tab:TB_params}), the difference in the $\gamma_\mathrm{X}$ parameters partly explains their MSD difference.
Thus, the here predicted higher h$^+$ mobility in \ce{MAPbBr3} compared to \ce{MAPbI3} (cf. Fig.~\ref{fig:avgMSD}, right panel), which agrees with experimentally observed trends \cite{Herz2017,Capitaine2021}, can in part be attributed to the stronger SOC in case of iodide.

\hspace{1em}
\section{\label{sec:level3}Conclusions}

In summary, this work investigated charge carrier transport in two prototypical HaP materials, namely \ce{MAPbBr3} and \ce{MAPbI3}. 
We used and further developed the DD model with a focus on the choice of electronic-structure theory and real-space charge transport channels.
Investigating temporal behaviors of orbital occupations, we examined the consequences of on-site energy levels and strength of SOC, which are determined by the chemical composition of the material.
We found that both on-site energies and the magnitude of SOC influence the relative orbital occupations and with that the dispersal of charge and carrier transport. 
Specifically, the energy gaps emerging across the on-site levels determine to which extent the charge-transport channels get filled over time. 
Identifying the $pp\pi$ channel as a bottleneck for charge transport, we could rationalize differences between e$^-$ and h$^+$ transport and explain different characteristics of the two investigated materials.
Altogether, the present study demonstrates the capabilities of a real-space analysis for the carrier dynamics in HaPs and motivates further development of related concepts for other material families.

\begin{acknowledgments}

Funding provided by Germany's Excellence Strategy – EXC 2089/1-390776260 , and by the Studienstiftung des Deutschen Volkes, are gratefully acknowledged. The authors further acknowledge the Gauss Centre for Supercomputing e.V. for funding this project by providing computing time through the John von Neumann Institute for Computing on the GCS Supercomputer JUWELS at Jülich Supercomputing Centre.

\end{acknowledgments}

\bibliographystyle{aipnum4-1}
\bibliography{reference}

\begin{thebibliography}{61}%
\makeatletter
\providecommand \@ifxundefined [1]{%
 \@ifx{#1\undefined}
}%
\providecommand \@ifnum [1]{%
 \ifnum #1\expandafter \@firstoftwo
 \else \expandafter \@secondoftwo
 \fi
}%
\providecommand \@ifx [1]{%
 \ifx #1\expandafter \@firstoftwo
 \else \expandafter \@secondoftwo
 \fi
}%
\providecommand \natexlab [1]{#1}%
\providecommand \enquote  [1]{``#1''}%
\providecommand \bibnamefont  [1]{#1}%
\providecommand \bibfnamefont [1]{#1}%
\providecommand \citenamefont [1]{#1}%
\providecommand \href@noop [0]{\@secondoftwo}%
\providecommand \href [0]{\begingroup \@sanitize@url \@href}%
\providecommand \@href[1]{\@@startlink{#1}\@@href}%
\providecommand \@@href[1]{\endgroup#1\@@endlink}%
\providecommand \@sanitize@url [0]{\catcode `\\12\catcode `\$12\catcode
  `\&12\catcode `\#12\catcode `\^12\catcode `\_12\catcode `\%12\relax}%
\providecommand \@@startlink[1]{}%
\providecommand \@@endlink[0]{}%
\providecommand \url  [0]{\begingroup\@sanitize@url \@url }%
\providecommand \@url [1]{\endgroup\@href {#1}{\urlprefix }}%
\providecommand \urlprefix  [0]{URL }%
\providecommand \Eprint [0]{\href }%
\providecommand \doibase [0]{http://dx.doi.org/}%
\providecommand \selectlanguage [0]{\@gobble}%
\providecommand \bibinfo  [0]{\@secondoftwo}%
\providecommand \bibfield  [0]{\@secondoftwo}%
\providecommand \translation [1]{[#1]}%
\providecommand \BibitemOpen [0]{}%
\providecommand \bibitemStop [0]{}%
\providecommand \bibitemNoStop [0]{.\EOS\space}%
\providecommand \EOS [0]{\spacefactor3000\relax}%
\providecommand \BibitemShut  [1]{\csname bibitem#1\endcsname}%
\let\auto@bib@innerbib\@empty
\bibitem [{\citenamefont {Stranks}\ and\ \citenamefont
  {Snaith}(2015)}]{review_devices}%
  \BibitemOpen
  \bibfield  {author} {\bibinfo {author} {\bibfnamefont {S.~D.}\ \bibnamefont
  {Stranks}}\ and\ \bibinfo {author} {\bibfnamefont {H.~J.}\ \bibnamefont
  {Snaith}},\ }\href {\doibase 10.1038/nnano.2015.90} {\bibfield  {journal}
  {\bibinfo  {journal} {Nat. Nanotechnol.}\ }\textbf {\bibinfo {volume} {10}},\
  \bibinfo {pages} {1748} (\bibinfo {year} {2015})}\BibitemShut {NoStop}%
\bibitem [{\citenamefont {Correa-Baena}\ \emph {et~al.}(2017)\citenamefont
  {Correa-Baena}, \citenamefont {Saliba}, \citenamefont {Buonassisi},
  \citenamefont {Grätzel}, \citenamefont {Abate}, \citenamefont {Tress},\ and\
  \citenamefont {Hagfeldt}}]{review_solarcell}%
  \BibitemOpen
  \bibfield  {author} {\bibinfo {author} {\bibfnamefont {J.-P.}\ \bibnamefont
  {Correa-Baena}}, \bibinfo {author} {\bibfnamefont {M.}~\bibnamefont
  {Saliba}}, \bibinfo {author} {\bibfnamefont {T.}~\bibnamefont {Buonassisi}},
  \bibinfo {author} {\bibfnamefont {M.}~\bibnamefont {Grätzel}}, \bibinfo
  {author} {\bibfnamefont {A.}~\bibnamefont {Abate}}, \bibinfo {author}
  {\bibfnamefont {W.}~\bibnamefont {Tress}}, \ and\ \bibinfo {author}
  {\bibfnamefont {A.}~\bibnamefont {Hagfeldt}},\ }\href
  {https://www.science.org/doi/abs/10.1126/science.aam6323} {\bibfield
  {journal} {\bibinfo  {journal} {Science}\ }\textbf {\bibinfo {volume}
  {358}},\ \bibinfo {pages} {739} (\bibinfo {year} {2017})}\BibitemShut
  {NoStop}%
\bibitem [{\citenamefont {Liu}\ \emph {et~al.}(2021)\citenamefont {Liu},
  \citenamefont {Qu}, \citenamefont {Kirchartz},\ and\ \citenamefont
  {Song}}]{review_opto_devices}%
  \BibitemOpen
  \bibfield  {author} {\bibinfo {author} {\bibfnamefont {J.}~\bibnamefont
  {Liu}}, \bibinfo {author} {\bibfnamefont {J.}~\bibnamefont {Qu}}, \bibinfo
  {author} {\bibfnamefont {T.}~\bibnamefont {Kirchartz}}, \ and\ \bibinfo
  {author} {\bibfnamefont {J.}~\bibnamefont {Song}},\ }\href {\doibase
  10.1039/D1TA04527J} {\bibfield  {journal} {\bibinfo  {journal} {J. Mater.
  Chem. A}\ }\textbf {\bibinfo {volume} {9}},\ \bibinfo {pages} {20919}
  (\bibinfo {year} {2021})}\BibitemShut {NoStop}%
\bibitem [{\citenamefont {Brenner}\ \emph {et~al.}(2016)\citenamefont
  {Brenner}, \citenamefont {Egger}, \citenamefont {Kronik}, \citenamefont
  {Hodes},\ and\ \citenamefont {Cahen}}]{transport_properties}%
  \BibitemOpen
  \bibfield  {author} {\bibinfo {author} {\bibfnamefont {T.~M.}\ \bibnamefont
  {Brenner}}, \bibinfo {author} {\bibfnamefont {D.~A.}\ \bibnamefont {Egger}},
  \bibinfo {author} {\bibfnamefont {L.}~\bibnamefont {Kronik}}, \bibinfo
  {author} {\bibfnamefont {G.}~\bibnamefont {Hodes}}, \ and\ \bibinfo {author}
  {\bibfnamefont {D.}~\bibnamefont {Cahen}},\ }\href {\doibase
  10.1038/natrevmats.2015.7} {\bibfield  {journal} {\bibinfo  {journal} {Nat.
  Rev. Mater.}\ }\textbf {\bibinfo {volume} {1}},\ \bibinfo {pages} {15007}
  (\bibinfo {year} {2016})}\BibitemShut {NoStop}%
\bibitem [{\citenamefont {Manser}, \citenamefont {Christians},\ and\
  \citenamefont {Kamat}(2016)}]{intrguing_optoelectronic}%
  \BibitemOpen
  \bibfield  {author} {\bibinfo {author} {\bibfnamefont {J.~S.}\ \bibnamefont
  {Manser}}, \bibinfo {author} {\bibfnamefont {J.~A.}\ \bibnamefont
  {Christians}}, \ and\ \bibinfo {author} {\bibfnamefont {P.~V.}\ \bibnamefont
  {Kamat}},\ }\href {https://doi.org/10.1021/acs.chemrev.6b00136} {\bibfield
  {journal} {\bibinfo  {journal} {Chem. Rev.}\ }\textbf {\bibinfo {volume}
  {116}},\ \bibinfo {pages} {12956} (\bibinfo {year} {2016})}\BibitemShut
  {NoStop}%
\bibitem [{\citenamefont {Stoumpos}\ and\ \citenamefont
  {Kanatzidis}(2016)}]{HaP_performance}%
  \BibitemOpen
  \bibfield  {author} {\bibinfo {author} {\bibfnamefont {C.~C.}\ \bibnamefont
  {Stoumpos}}\ and\ \bibinfo {author} {\bibfnamefont {M.~G.}\ \bibnamefont
  {Kanatzidis}},\ }\href {\doibase https://doi.org/10.1002/adma.201600265}
  {\bibfield  {journal} {\bibinfo  {journal} {Adv. Mater.}\ }\textbf {\bibinfo
  {volume} {28}},\ \bibinfo {pages} {5778} (\bibinfo {year}
  {2016})}\BibitemShut {NoStop}%
\bibitem [{\citenamefont {Li}\ \emph {et~al.}(2017)\citenamefont {Li},
  \citenamefont {Wang}, \citenamefont {Deschler}, \citenamefont {Gao},
  \citenamefont {Friend},\ and\ \citenamefont
  {Cheetham}}]{review_HaP_properties}%
  \BibitemOpen
  \bibfield  {author} {\bibinfo {author} {\bibfnamefont {W.}~\bibnamefont
  {Li}}, \bibinfo {author} {\bibfnamefont {Z.}~\bibnamefont {Wang}}, \bibinfo
  {author} {\bibfnamefont {F.}~\bibnamefont {Deschler}}, \bibinfo {author}
  {\bibfnamefont {S.}~\bibnamefont {Gao}}, \bibinfo {author} {\bibfnamefont
  {R.~H.}\ \bibnamefont {Friend}}, \ and\ \bibinfo {author} {\bibfnamefont
  {A.~K.}\ \bibnamefont {Cheetham}},\ }\href {\doibase
  10.1038/natrevmats.2016.99} {\bibfield  {journal} {\bibinfo  {journal} {Nat.
  Rev. Mater.}\ }\textbf {\bibinfo {volume} {2}},\ \bibinfo {pages} {2058}
  (\bibinfo {year} {2017})}\BibitemShut {NoStop}%
\bibitem [{\citenamefont {De~Wolf}\ \emph {et~al.}(2014)\citenamefont
  {De~Wolf}, \citenamefont {Holovsky}, \citenamefont {Moon}, \citenamefont
  {Löper}, \citenamefont {Niesen}, \citenamefont {Ledinsky}, \citenamefont
  {Haug}, \citenamefont {Yum},\ and\ \citenamefont {Ballif}}]{absorption}%
  \BibitemOpen
  \bibfield  {author} {\bibinfo {author} {\bibfnamefont {S.}~\bibnamefont
  {De~Wolf}}, \bibinfo {author} {\bibfnamefont {J.}~\bibnamefont {Holovsky}},
  \bibinfo {author} {\bibfnamefont {S.-J.}\ \bibnamefont {Moon}}, \bibinfo
  {author} {\bibfnamefont {P.}~\bibnamefont {Löper}}, \bibinfo {author}
  {\bibfnamefont {B.}~\bibnamefont {Niesen}}, \bibinfo {author} {\bibfnamefont
  {M.}~\bibnamefont {Ledinsky}}, \bibinfo {author} {\bibfnamefont {F.-J.}\
  \bibnamefont {Haug}}, \bibinfo {author} {\bibfnamefont {J.-H.}\ \bibnamefont
  {Yum}}, \ and\ \bibinfo {author} {\bibfnamefont {C.}~\bibnamefont {Ballif}},\
  }\href {\doibase 10.1021/jz500279b} {\bibfield  {journal} {\bibinfo
  {journal} {J. Phys. Chem. Lett.}\ }\textbf {\bibinfo {volume} {5}},\ \bibinfo
  {pages} {1035} (\bibinfo {year} {2014})}\BibitemShut {NoStop}%
\bibitem [{\citenamefont {Baranowski}\ and\ \citenamefont
  {Plochocka}(2020)}]{exciton_binding}%
  \BibitemOpen
  \bibfield  {author} {\bibinfo {author} {\bibfnamefont {M.}~\bibnamefont
  {Baranowski}}\ and\ \bibinfo {author} {\bibfnamefont {P.}~\bibnamefont
  {Plochocka}},\ }\href {\doibase https://doi.org/10.1002/aenm.201903659}
  {\bibfield  {journal} {\bibinfo  {journal} {Adv. Energy Mater.}\ }\textbf
  {\bibinfo {volume} {10}},\ \bibinfo {pages} {1903659} (\bibinfo {year}
  {2020})}\BibitemShut {NoStop}%
\bibitem [{\citenamefont {Stranks}\ \emph {et~al.}(2013)\citenamefont
  {Stranks}, \citenamefont {Eperon}, \citenamefont {Grancini}, \citenamefont
  {Menelaou}, \citenamefont {Alcocer}, \citenamefont {Leijtens}, \citenamefont
  {Herz}, \citenamefont {Petrozza},\ and\ \citenamefont
  {Snaith}}]{diffusion_length2}%
  \BibitemOpen
  \bibfield  {author} {\bibinfo {author} {\bibfnamefont {S.~D.}\ \bibnamefont
  {Stranks}}, \bibinfo {author} {\bibfnamefont {G.~E.}\ \bibnamefont {Eperon}},
  \bibinfo {author} {\bibfnamefont {G.}~\bibnamefont {Grancini}}, \bibinfo
  {author} {\bibfnamefont {C.}~\bibnamefont {Menelaou}}, \bibinfo {author}
  {\bibfnamefont {M.~J.~P.}\ \bibnamefont {Alcocer}}, \bibinfo {author}
  {\bibfnamefont {T.}~\bibnamefont {Leijtens}}, \bibinfo {author}
  {\bibfnamefont {L.~M.}\ \bibnamefont {Herz}}, \bibinfo {author}
  {\bibfnamefont {A.}~\bibnamefont {Petrozza}}, \ and\ \bibinfo {author}
  {\bibfnamefont {H.~J.}\ \bibnamefont {Snaith}},\ }\href {\doibase
  10.1126/science.1243982} {\bibfield  {journal} {\bibinfo  {journal}
  {Science}\ }\textbf {\bibinfo {volume} {342}},\ \bibinfo {pages} {341}
  (\bibinfo {year} {2013})}\BibitemShut {NoStop}%
\bibitem [{\citenamefont {Johnston}\ and\ \citenamefont
  {Herz}(2016)}]{diffusion_length}%
  \BibitemOpen
  \bibfield  {author} {\bibinfo {author} {\bibfnamefont {M.~B.}\ \bibnamefont
  {Johnston}}\ and\ \bibinfo {author} {\bibfnamefont {L.~M.}\ \bibnamefont
  {Herz}},\ }\href {\doibase 10.1021/acs.accounts.5b00411} {\bibfield
  {journal} {\bibinfo  {journal} {Acc. Chem. Res.}\ }\textbf {\bibinfo {volume}
  {49}},\ \bibinfo {pages} {146} (\bibinfo {year} {2016})}\BibitemShut
  {NoStop}%
\bibitem [{\citenamefont {Unger}\ \emph {et~al.}(2017)\citenamefont {Unger},
  \citenamefont {Kegelmann}, \citenamefont {Suchan}, \citenamefont {Sörell},
  \citenamefont {Korte},\ and\ \citenamefont {Albrecht}}]{bandgap_tuning}%
  \BibitemOpen
  \bibfield  {author} {\bibinfo {author} {\bibfnamefont {E.~L.}\ \bibnamefont
  {Unger}}, \bibinfo {author} {\bibfnamefont {L.}~\bibnamefont {Kegelmann}},
  \bibinfo {author} {\bibfnamefont {K.}~\bibnamefont {Suchan}}, \bibinfo
  {author} {\bibfnamefont {D.}~\bibnamefont {Sörell}}, \bibinfo {author}
  {\bibfnamefont {L.}~\bibnamefont {Korte}}, \ and\ \bibinfo {author}
  {\bibfnamefont {S.}~\bibnamefont {Albrecht}},\ }\href {\doibase
  10.1039/C7TA00404D} {\bibfield  {journal} {\bibinfo  {journal} {J. Mater.
  Chem. A}\ }\textbf {\bibinfo {volume} {5}},\ \bibinfo {pages} {11401}
  (\bibinfo {year} {2017})}\BibitemShut {NoStop}%
\bibitem [{\citenamefont {Yu}\ and\ \citenamefont
  {Cardona}(2010)}]{Cardona_Yu}%
  \BibitemOpen
  \bibfield  {author} {\bibinfo {author} {\bibfnamefont {P.~Y.}\ \bibnamefont
  {Yu}}\ and\ \bibinfo {author} {\bibfnamefont {M.}~\bibnamefont {Cardona}},\
  }\href {\doibase 10.1007/978-3-642-00710-1} {\emph {\bibinfo {title}
  {Fundamentals of Semiconductors}}}\ (\bibinfo  {publisher} {Springer-Verlag
  Berlin Heidelberg},\ \bibinfo {year} {2010})\BibitemShut {NoStop}%
\bibitem [{\citenamefont {Schilcher}\ \emph {et~al.}(2021)\citenamefont
  {Schilcher}, \citenamefont {Robinson}, \citenamefont {Abramovitch},
  \citenamefont {Tan}, \citenamefont {Rappe}, \citenamefont {Reichman},\ and\
  \citenamefont {Egger}}]{MaxPaper1}%
  \BibitemOpen
  \bibfield  {author} {\bibinfo {author} {\bibfnamefont {M.~J.}\ \bibnamefont
  {Schilcher}}, \bibinfo {author} {\bibfnamefont {P.~J.}\ \bibnamefont
  {Robinson}}, \bibinfo {author} {\bibfnamefont {D.~J.}\ \bibnamefont
  {Abramovitch}}, \bibinfo {author} {\bibfnamefont {L.~Z.}\ \bibnamefont
  {Tan}}, \bibinfo {author} {\bibfnamefont {A.~M.}\ \bibnamefont {Rappe}},
  \bibinfo {author} {\bibfnamefont {D.~R.}\ \bibnamefont {Reichman}}, \ and\
  \bibinfo {author} {\bibfnamefont {D.~A.}\ \bibnamefont {Egger}},\ }\href
  {\doibase 10.1021/acsenergylett.1c00506} {\bibfield  {journal} {\bibinfo
  {journal} {ACS Energy Lett.}\ }\textbf {\bibinfo {volume} {6}},\ \bibinfo
  {pages} {2162} (\bibinfo {year} {2021})}\BibitemShut {NoStop}%
\bibitem [{\citenamefont {Troisi}\ and\ \citenamefont
  {Orlandi}(2006)}]{Troisi2006}%
  \BibitemOpen
  \bibfield  {author} {\bibinfo {author} {\bibfnamefont {A.}~\bibnamefont
  {Troisi}}\ and\ \bibinfo {author} {\bibfnamefont {G.}~\bibnamefont
  {Orlandi}},\ }\href {\doibase 10.1103/PhysRevLett.96.086601} {\bibfield
  {journal} {\bibinfo  {journal} {Phys. Rev. Lett.}\ }\textbf {\bibinfo
  {volume} {96}},\ \bibinfo {pages} {086601} (\bibinfo {year}
  {2006})}\BibitemShut {NoStop}%
\bibitem [{\citenamefont {Fratini}\ and\ \citenamefont
  {Ciuchi}(2009)}]{Fratini2009}%
  \BibitemOpen
  \bibfield  {author} {\bibinfo {author} {\bibfnamefont {S.}~\bibnamefont
  {Fratini}}\ and\ \bibinfo {author} {\bibfnamefont {S.}~\bibnamefont
  {Ciuchi}},\ }\href {\doibase 10.1103/PhysRevLett.103.266601} {\bibfield
  {journal} {\bibinfo  {journal} {Phys. Rev. Lett.}\ }\textbf {\bibinfo
  {volume} {103}},\ \bibinfo {pages} {266601} (\bibinfo {year}
  {2009})}\BibitemShut {NoStop}%
\bibitem [{\citenamefont {Mishchenko}\ \emph {et~al.}(2019)\citenamefont
  {Mishchenko}, \citenamefont {Pollet}, \citenamefont {Prokof'ev},
  \citenamefont {Kumar}, \citenamefont {Maslov},\ and\ \citenamefont
  {Nagaosa}}]{beyond_polaron}%
  \BibitemOpen
  \bibfield  {author} {\bibinfo {author} {\bibfnamefont {A.~S.}\ \bibnamefont
  {Mishchenko}}, \bibinfo {author} {\bibfnamefont {L.}~\bibnamefont {Pollet}},
  \bibinfo {author} {\bibfnamefont {N.~V.}\ \bibnamefont {Prokof'ev}}, \bibinfo
  {author} {\bibfnamefont {A.}~\bibnamefont {Kumar}}, \bibinfo {author}
  {\bibfnamefont {D.~L.}\ \bibnamefont {Maslov}}, \ and\ \bibinfo {author}
  {\bibfnamefont {N.}~\bibnamefont {Nagaosa}},\ }\href
  {https://link.aps.org/doi/10.1103/PhysRevLett.123.076601} {\bibfield
  {journal} {\bibinfo  {journal} {Phys. Rev. Lett.}\ }\textbf {\bibinfo
  {volume} {123}},\ \bibinfo {pages} {076601} (\bibinfo {year}
  {2019})}\BibitemShut {NoStop}%
\bibitem [{\citenamefont {Giannini}\ \emph {et~al.}(2020)\citenamefont
  {Giannini}, \citenamefont {Ziogos}, \citenamefont {Carof}, \citenamefont
  {Ellis},\ and\ \citenamefont {Blumberger}}]{Blumberger2020}%
  \BibitemOpen
  \bibfield  {author} {\bibinfo {author} {\bibfnamefont {S.}~\bibnamefont
  {Giannini}}, \bibinfo {author} {\bibfnamefont {O.~G.}\ \bibnamefont
  {Ziogos}}, \bibinfo {author} {\bibfnamefont {A.}~\bibnamefont {Carof}},
  \bibinfo {author} {\bibfnamefont {M.}~\bibnamefont {Ellis}}, \ and\ \bibinfo
  {author} {\bibfnamefont {J.}~\bibnamefont {Blumberger}},\ }\href {\doibase
  https://doi.org/10.1002/adts.202000093} {\bibfield  {journal} {\bibinfo
  {journal} {Adv. Theory Simul.}\ }\textbf {\bibinfo {volume} {3}},\ \bibinfo
  {pages} {2000093} (\bibinfo {year} {2020})}\BibitemShut {NoStop}%
\bibitem [{\citenamefont {Quarti}\ \emph {et~al.}(2016)\citenamefont {Quarti},
  \citenamefont {Mosconi}, \citenamefont {Ball}, \citenamefont {D{'}Innocenzo},
  \citenamefont {Tao}, \citenamefont {Pathak}, \citenamefont {Snaith},
  \citenamefont {Petrozza},\ and\ \citenamefont {De~Angelis}}]{Quarti2016}%
  \BibitemOpen
  \bibfield  {author} {\bibinfo {author} {\bibfnamefont {C.}~\bibnamefont
  {Quarti}}, \bibinfo {author} {\bibfnamefont {E.}~\bibnamefont {Mosconi}},
  \bibinfo {author} {\bibfnamefont {J.~M.}\ \bibnamefont {Ball}}, \bibinfo
  {author} {\bibfnamefont {V.}~\bibnamefont {D{'}Innocenzo}}, \bibinfo {author}
  {\bibfnamefont {C.}~\bibnamefont {Tao}}, \bibinfo {author} {\bibfnamefont
  {S.}~\bibnamefont {Pathak}}, \bibinfo {author} {\bibfnamefont {H.~J.}\
  \bibnamefont {Snaith}}, \bibinfo {author} {\bibfnamefont {A.}~\bibnamefont
  {Petrozza}}, \ and\ \bibinfo {author} {\bibfnamefont {F.}~\bibnamefont
  {De~Angelis}},\ }\href {\doibase 10.1039/C5EE02925B} {\bibfield  {journal}
  {\bibinfo  {journal} {Energy Environ. Sci.}\ }\textbf {\bibinfo {volume}
  {9}},\ \bibinfo {pages} {155} (\bibinfo {year} {2016})}\BibitemShut {NoStop}%
\bibitem [{\citenamefont {Yaffe}\ \emph {et~al.}(2017)\citenamefont {Yaffe},
  \citenamefont {Guo}, \citenamefont {Tan}, \citenamefont {Egger},
  \citenamefont {Hull}, \citenamefont {Stoumpos}, \citenamefont {Zheng},
  \citenamefont {Heinz}, \citenamefont {Kronik}, \citenamefont {Kanatzidis},
  \citenamefont {Owen}, \citenamefont {Rappe}, \citenamefont {Pimenta},\ and\
  \citenamefont {Brus}}]{Yaffe2017}%
  \BibitemOpen
  \bibfield  {author} {\bibinfo {author} {\bibfnamefont {O.}~\bibnamefont
  {Yaffe}}, \bibinfo {author} {\bibfnamefont {Y.}~\bibnamefont {Guo}}, \bibinfo
  {author} {\bibfnamefont {L.~Z.}\ \bibnamefont {Tan}}, \bibinfo {author}
  {\bibfnamefont {D.~A.}\ \bibnamefont {Egger}}, \bibinfo {author}
  {\bibfnamefont {T.}~\bibnamefont {Hull}}, \bibinfo {author} {\bibfnamefont
  {C.~C.}\ \bibnamefont {Stoumpos}}, \bibinfo {author} {\bibfnamefont
  {F.}~\bibnamefont {Zheng}}, \bibinfo {author} {\bibfnamefont {T.~F.}\
  \bibnamefont {Heinz}}, \bibinfo {author} {\bibfnamefont {L.}~\bibnamefont
  {Kronik}}, \bibinfo {author} {\bibfnamefont {M.~G.}\ \bibnamefont
  {Kanatzidis}}, \bibinfo {author} {\bibfnamefont {J.~S.}\ \bibnamefont
  {Owen}}, \bibinfo {author} {\bibfnamefont {A.~M.}\ \bibnamefont {Rappe}},
  \bibinfo {author} {\bibfnamefont {M.~A.}\ \bibnamefont {Pimenta}}, \ and\
  \bibinfo {author} {\bibfnamefont {L.~E.}\ \bibnamefont {Brus}},\ }\href
  {\doibase 10.1103/PhysRevLett.118.136001} {\bibfield  {journal} {\bibinfo
  {journal} {Phys. Rev. Lett.}\ }\textbf {\bibinfo {volume} {118}},\ \bibinfo
  {pages} {136001} (\bibinfo {year} {2017})}\BibitemShut {NoStop}%
\bibitem [{\citenamefont {Marronnier}\ \emph {et~al.}(2018)\citenamefont
  {Marronnier}, \citenamefont {Roma}, \citenamefont {Boyer-Richard},
  \citenamefont {Pedesseau}, \citenamefont {Jancu}, \citenamefont
  {Bonnassieux}, \citenamefont {Katan}, \citenamefont {Stoumpos}, \citenamefont
  {Kanatzidis},\ and\ \citenamefont {Even}}]{Marronnier2018}%
  \BibitemOpen
  \bibfield  {author} {\bibinfo {author} {\bibfnamefont {A.}~\bibnamefont
  {Marronnier}}, \bibinfo {author} {\bibfnamefont {G.}~\bibnamefont {Roma}},
  \bibinfo {author} {\bibfnamefont {S.}~\bibnamefont {Boyer-Richard}}, \bibinfo
  {author} {\bibfnamefont {L.}~\bibnamefont {Pedesseau}}, \bibinfo {author}
  {\bibfnamefont {J.-M.}\ \bibnamefont {Jancu}}, \bibinfo {author}
  {\bibfnamefont {Y.}~\bibnamefont {Bonnassieux}}, \bibinfo {author}
  {\bibfnamefont {C.}~\bibnamefont {Katan}}, \bibinfo {author} {\bibfnamefont
  {C.~C.}\ \bibnamefont {Stoumpos}}, \bibinfo {author} {\bibfnamefont {M.~G.}\
  \bibnamefont {Kanatzidis}}, \ and\ \bibinfo {author} {\bibfnamefont
  {J.}~\bibnamefont {Even}},\ }\href {\doibase 10.1021/acsnano.8b00267}
  {\bibfield  {journal} {\bibinfo  {journal} {ACS Nano}\ }\textbf {\bibinfo
  {volume} {12}},\ \bibinfo {pages} {3477} (\bibinfo {year}
  {2018})}\BibitemShut {NoStop}%
\bibitem [{\citenamefont {Mayers}\ \emph {et~al.}(2018)\citenamefont {Mayers},
  \citenamefont {Tan}, \citenamefont {Egger}, \citenamefont {Rappe},\ and\
  \citenamefont {Reichman}}]{Mayers}%
  \BibitemOpen
  \bibfield  {author} {\bibinfo {author} {\bibfnamefont {M.~Z.}\ \bibnamefont
  {Mayers}}, \bibinfo {author} {\bibfnamefont {L.~Z.}\ \bibnamefont {Tan}},
  \bibinfo {author} {\bibfnamefont {D.~A.}\ \bibnamefont {Egger}}, \bibinfo
  {author} {\bibfnamefont {A.~M.}\ \bibnamefont {Rappe}}, \ and\ \bibinfo
  {author} {\bibfnamefont {D.~R.}\ \bibnamefont {Reichman}},\ }\href
  {https://doi.org/10.1021/acs.nanolett.8b04276} {\bibfield  {journal}
  {\bibinfo  {journal} {Nano Lett.}\ }\textbf {\bibinfo {volume} {18}},\
  \bibinfo {pages} {8041} (\bibinfo {year} {2018})}\BibitemShut {NoStop}%
\bibitem [{\citenamefont {Gehrmann}\ and\ \citenamefont
  {Egger}(2019)}]{Gehrmann2019}%
  \BibitemOpen
  \bibfield  {author} {\bibinfo {author} {\bibfnamefont {C.}~\bibnamefont
  {Gehrmann}}\ and\ \bibinfo {author} {\bibfnamefont {D.~A.}\ \bibnamefont
  {Egger}},\ }\href {\doibase 10.1038/s41467-019-11087-y} {\bibfield  {journal}
  {\bibinfo  {journal} {Nat. Commun.}\ }\textbf {\bibinfo {volume} {10}},\
  \bibinfo {pages} {3141} (\bibinfo {year} {2019})}\BibitemShut {NoStop}%
\bibitem [{\citenamefont {Schilcher}\ \emph {et~al.}(2023)\citenamefont
  {Schilcher}, \citenamefont {Abramovitch}, \citenamefont {Mayers},
  \citenamefont {Tan}, \citenamefont {Reichman},\ and\ \citenamefont
  {Egger}}]{MaxPaper2}%
  \BibitemOpen
  \bibfield  {author} {\bibinfo {author} {\bibfnamefont {M.~J.}\ \bibnamefont
  {Schilcher}}, \bibinfo {author} {\bibfnamefont {D.~J.}\ \bibnamefont
  {Abramovitch}}, \bibinfo {author} {\bibfnamefont {M.~Z.}\ \bibnamefont
  {Mayers}}, \bibinfo {author} {\bibfnamefont {L.~Z.}\ \bibnamefont {Tan}},
  \bibinfo {author} {\bibfnamefont {D.~R.}\ \bibnamefont {Reichman}}, \ and\
  \bibinfo {author} {\bibfnamefont {D.~A.}\ \bibnamefont {Egger}},\ }\href
  {\doibase 10.1103/PhysRevMaterials.7.L081601} {\bibfield  {journal} {\bibinfo
   {journal} {Phys. Rev. Mater.}\ }\textbf {\bibinfo {volume} {7}},\ \bibinfo
  {pages} {L081601} (\bibinfo {year} {2023})}\BibitemShut {NoStop}%
\bibitem [{\citenamefont {Seidl}\ \emph {et~al.}(2023)\citenamefont {Seidl},
  \citenamefont {Zhu}, \citenamefont {Reuveni}, \citenamefont {Aharon},
  \citenamefont {Gehrmann}, \citenamefont {Caicedo-D\'avila}, \citenamefont
  {Yaffe},\ and\ \citenamefont {Egger}}]{Seidl2023}%
  \BibitemOpen
  \bibfield  {author} {\bibinfo {author} {\bibfnamefont {S.~A.}\ \bibnamefont
  {Seidl}}, \bibinfo {author} {\bibfnamefont {X.}~\bibnamefont {Zhu}}, \bibinfo
  {author} {\bibfnamefont {G.}~\bibnamefont {Reuveni}}, \bibinfo {author}
  {\bibfnamefont {S.}~\bibnamefont {Aharon}}, \bibinfo {author} {\bibfnamefont
  {C.}~\bibnamefont {Gehrmann}}, \bibinfo {author} {\bibfnamefont
  {S.}~\bibnamefont {Caicedo-D\'avila}}, \bibinfo {author} {\bibfnamefont
  {O.}~\bibnamefont {Yaffe}}, \ and\ \bibinfo {author} {\bibfnamefont {D.~A.}\
  \bibnamefont {Egger}},\ }\href {\doibase 10.1103/PhysRevMaterials.7.L092401}
  {\bibfield  {journal} {\bibinfo  {journal} {Phys. Rev. Mater.}\ }\textbf
  {\bibinfo {volume} {7}},\ \bibinfo {pages} {L092401} (\bibinfo {year}
  {2023})}\BibitemShut {NoStop}%
\bibitem [{\citenamefont {Zacharias}\ \emph {et~al.}(2023)\citenamefont
  {Zacharias}, \citenamefont {Volonakis}, \citenamefont {Giustino},\ and\
  \citenamefont {Even}}]{Zacharias2023}%
  \BibitemOpen
  \bibfield  {author} {\bibinfo {author} {\bibfnamefont {M.}~\bibnamefont
  {Zacharias}}, \bibinfo {author} {\bibfnamefont {G.}~\bibnamefont
  {Volonakis}}, \bibinfo {author} {\bibfnamefont {F.}~\bibnamefont {Giustino}},
  \ and\ \bibinfo {author} {\bibfnamefont {J.}~\bibnamefont {Even}},\ }\href
  {\doibase 10.1038/s41524-023-01089-2} {\bibfield  {journal} {\bibinfo
  {journal} {Npj Comput. Mater.}\ }\textbf {\bibinfo {volume} {9}},\ \bibinfo
  {pages} {153} (\bibinfo {year} {2023})}\BibitemShut {NoStop}%
\bibitem [{\citenamefont {Frenzel}\ \emph {et~al.}(2023)\citenamefont
  {Frenzel}, \citenamefont {Cherasse}, \citenamefont {Urban}, \citenamefont
  {Wang}, \citenamefont {Xiang}, \citenamefont {Nest}, \citenamefont {Huber},
  \citenamefont {Perfetti}, \citenamefont {Wolf}, \citenamefont {Kampfrath},
  \citenamefont {Zhu},\ and\ \citenamefont {Maehrlein}}]{Frenzel2023}%
  \BibitemOpen
  \bibfield  {author} {\bibinfo {author} {\bibfnamefont {M.}~\bibnamefont
  {Frenzel}}, \bibinfo {author} {\bibfnamefont {M.}~\bibnamefont {Cherasse}},
  \bibinfo {author} {\bibfnamefont {J.~M.}\ \bibnamefont {Urban}}, \bibinfo
  {author} {\bibfnamefont {F.}~\bibnamefont {Wang}}, \bibinfo {author}
  {\bibfnamefont {B.}~\bibnamefont {Xiang}}, \bibinfo {author} {\bibfnamefont
  {L.}~\bibnamefont {Nest}}, \bibinfo {author} {\bibfnamefont {L.}~\bibnamefont
  {Huber}}, \bibinfo {author} {\bibfnamefont {L.}~\bibnamefont {Perfetti}},
  \bibinfo {author} {\bibfnamefont {M.}~\bibnamefont {Wolf}}, \bibinfo {author}
  {\bibfnamefont {T.}~\bibnamefont {Kampfrath}}, \bibinfo {author}
  {\bibfnamefont {X.-Y.}\ \bibnamefont {Zhu}}, \ and\ \bibinfo {author}
  {\bibfnamefont {S.~F.}\ \bibnamefont {Maehrlein}},\ }\href
  {https://www.science.org/doi/abs/10.1126/sciadv.adg3856} {\bibfield
  {journal} {\bibinfo  {journal} {Sci. Adv.}\ }\textbf {\bibinfo {volume}
  {9}},\ \bibinfo {pages} {eadg3856} (\bibinfo {year} {2023})}\BibitemShut
  {NoStop}%
\bibitem [{\citenamefont {Park}\ and\ \citenamefont {Limmer}(2023)}]{Park2023}%
  \BibitemOpen
  \bibfield  {author} {\bibinfo {author} {\bibfnamefont {Y.}~\bibnamefont
  {Park}}\ and\ \bibinfo {author} {\bibfnamefont {D.~T.}\ \bibnamefont
  {Limmer}},\ }\href {\doibase 10.1103/PhysRevMaterials.7.106002} {\bibfield
  {journal} {\bibinfo  {journal} {Phys. Rev. Mater.}\ }\textbf {\bibinfo
  {volume} {7}},\ \bibinfo {pages} {106002} (\bibinfo {year}
  {2023})}\BibitemShut {NoStop}%
\bibitem [{\citenamefont {Wiktor}\ \emph {et~al.}(2023)\citenamefont {Wiktor},
  \citenamefont {Fransson}, \citenamefont {Kubicki},\ and\ \citenamefont
  {Erhart}}]{Wiktor2023}%
  \BibitemOpen
  \bibfield  {author} {\bibinfo {author} {\bibfnamefont {J.}~\bibnamefont
  {Wiktor}}, \bibinfo {author} {\bibfnamefont {E.}~\bibnamefont {Fransson}},
  \bibinfo {author} {\bibfnamefont {D.}~\bibnamefont {Kubicki}}, \ and\
  \bibinfo {author} {\bibfnamefont {P.}~\bibnamefont {Erhart}},\ }\href
  {\doibase 10.1021/acs.chemmater.3c00933} {\bibfield  {journal} {\bibinfo
  {journal} {Chem. Mat.}\ }\textbf {\bibinfo {volume} {35}},\ \bibinfo {pages}
  {6737} (\bibinfo {year} {2023})}\BibitemShut {NoStop}%
\bibitem [{\citenamefont {Hylton-Farrington}\ and\ \citenamefont
  {Remsing}(2024)}]{HyltonFarrington2024}%
  \BibitemOpen
  \bibfield  {author} {\bibinfo {author} {\bibfnamefont {C.~M.}\ \bibnamefont
  {Hylton-Farrington}}\ and\ \bibinfo {author} {\bibfnamefont {R.~C.}\
  \bibnamefont {Remsing}},\ }\href {\doibase 10.1021/acs.chemmater.4c01121}
  {\bibfield  {journal} {\bibinfo  {journal} {Chem. Mater.}\ }\textbf {\bibinfo
  {volume} {36}},\ \bibinfo {pages} {9442} (\bibinfo {year}
  {2024})}\BibitemShut {NoStop}%
\bibitem [{\citenamefont {Zhu}\ and\ \citenamefont {Egger}(2025)}]{Zhu2025}%
  \BibitemOpen
  \bibfield  {author} {\bibinfo {author} {\bibfnamefont {X.}~\bibnamefont
  {Zhu}}\ and\ \bibinfo {author} {\bibfnamefont {D.~A.}\ \bibnamefont
  {Egger}},\ }\href {\doibase 10.1103/PhysRevLett.134.016403} {\bibfield
  {journal} {\bibinfo  {journal} {Phys. Rev. Lett.}\ }\textbf {\bibinfo
  {volume} {134}},\ \bibinfo {pages} {016403} (\bibinfo {year}
  {2025})}\BibitemShut {NoStop}%
\bibitem [{\citenamefont {Ioffe}\ and\ \citenamefont
  {Regel}(1960)}]{Ioffe_Regel}%
  \BibitemOpen
  \bibfield  {author} {\bibinfo {author} {\bibfnamefont {A.~F.}\ \bibnamefont
  {Ioffe}}\ and\ \bibinfo {author} {\bibfnamefont {A.~R.}\ \bibnamefont
  {Regel}},\ }\href@noop {} {\bibfield  {journal} {\bibinfo  {journal} {Prog.
  Semicond}\ }\textbf {\bibinfo {volume} {4}},\ \bibinfo {pages} {237}
  (\bibinfo {year} {1960})}\BibitemShut {NoStop}%
\bibitem [{\citenamefont {Mott}(1990)}]{Mott}%
  \BibitemOpen
  \bibfield  {author} {\bibinfo {author} {\bibfnamefont {N.}~\bibnamefont
  {Mott}},\ }\href {\doibase https://doi.org/10.1201/b12795} {\emph {\bibinfo
  {title} {Metal-Insulator Tranisitions}}}\ (\bibinfo  {publisher} {CRC
  Press.},\ \bibinfo {year} {1990})\BibitemShut {NoStop}%
\bibitem [{\citenamefont {Wiktor}\ \emph {et~al.}(2017)\citenamefont {Wiktor},
  \citenamefont {Reshetnyak}, \citenamefont {Ambrosio},\ and\ \citenamefont
  {Pasquarello}}]{Wiktor2017}%
  \BibitemOpen
  \bibfield  {author} {\bibinfo {author} {\bibfnamefont {J.}~\bibnamefont
  {Wiktor}}, \bibinfo {author} {\bibfnamefont {I.}~\bibnamefont {Reshetnyak}},
  \bibinfo {author} {\bibfnamefont {F.}~\bibnamefont {Ambrosio}}, \ and\
  \bibinfo {author} {\bibfnamefont {A.}~\bibnamefont {Pasquarello}},\ }\href
  {\doibase 10.1103/PhysRevMaterials.1.022401} {\bibfield  {journal} {\bibinfo
  {journal} {Phys. Rev. Mater.}\ }\textbf {\bibinfo {volume} {1}},\ \bibinfo
  {pages} {022401} (\bibinfo {year} {2017})}\BibitemShut {NoStop}%
\bibitem [{\citenamefont {Zhou}\ and\ \citenamefont {Bernardi}(2019)}]{SrTiO3}%
  \BibitemOpen
  \bibfield  {author} {\bibinfo {author} {\bibfnamefont {J.-J.}\ \bibnamefont
  {Zhou}}\ and\ \bibinfo {author} {\bibfnamefont {M.}~\bibnamefont
  {Bernardi}},\ }\href {\doibase 10.1103/PhysRevResearch.1.033138} {\bibfield
  {journal} {\bibinfo  {journal} {Phys. Rev. Res.}\ }\textbf {\bibinfo {volume}
  {1}},\ \bibinfo {pages} {033138} (\bibinfo {year} {2019})}\BibitemShut
  {NoStop}%
\bibitem [{\citenamefont {Hegner}\ \emph {et~al.}(2024)\citenamefont {Hegner},
  \citenamefont {Cohen}, \citenamefont {Rudel}, \citenamefont {Kronawitter},
  \citenamefont {Grumet}, \citenamefont {Zhu}, \citenamefont {Korobko},
  \citenamefont {Houben}, \citenamefont {Jiang}, \citenamefont {Schnick},
  \citenamefont {Kieslich}, \citenamefont {Yaffe}, \citenamefont {Sharp},\ and\
  \citenamefont {Egger}}]{FranziskaPaper}%
  \BibitemOpen
  \bibfield  {author} {\bibinfo {author} {\bibfnamefont {F.~S.}\ \bibnamefont
  {Hegner}}, \bibinfo {author} {\bibfnamefont {A.}~\bibnamefont {Cohen}},
  \bibinfo {author} {\bibfnamefont {S.~S.}\ \bibnamefont {Rudel}}, \bibinfo
  {author} {\bibfnamefont {S.~M.}\ \bibnamefont {Kronawitter}}, \bibinfo
  {author} {\bibfnamefont {M.}~\bibnamefont {Grumet}}, \bibinfo {author}
  {\bibfnamefont {X.}~\bibnamefont {Zhu}}, \bibinfo {author} {\bibfnamefont
  {R.}~\bibnamefont {Korobko}}, \bibinfo {author} {\bibfnamefont
  {L.}~\bibnamefont {Houben}}, \bibinfo {author} {\bibfnamefont {C.-M.}\
  \bibnamefont {Jiang}}, \bibinfo {author} {\bibfnamefont {W.}~\bibnamefont
  {Schnick}}, \bibinfo {author} {\bibfnamefont {G.}~\bibnamefont {Kieslich}},
  \bibinfo {author} {\bibfnamefont {O.}~\bibnamefont {Yaffe}}, \bibinfo
  {author} {\bibfnamefont {I.~D.}\ \bibnamefont {Sharp}}, \ and\ \bibinfo
  {author} {\bibfnamefont {D.~A.}\ \bibnamefont {Egger}},\ }\href {\doibase
  https://doi.org/10.1002/aenm.202303059} {\bibfield  {journal} {\bibinfo
  {journal} {Adv. Energy Mater.}\ }\textbf {\bibinfo {volume} {14}},\ \bibinfo
  {pages} {2303059} (\bibinfo {year} {2024})}\BibitemShut {NoStop}%
\bibitem [{\citenamefont {Wang}\ \emph {et~al.}(2011)\citenamefont {Wang},
  \citenamefont {Beljonne}, \citenamefont {Chen},\ and\ \citenamefont
  {Shi}}]{Beljonne2011}%
  \BibitemOpen
  \bibfield  {author} {\bibinfo {author} {\bibfnamefont {L.}~\bibnamefont
  {Wang}}, \bibinfo {author} {\bibfnamefont {D.}~\bibnamefont {Beljonne}},
  \bibinfo {author} {\bibfnamefont {L.}~\bibnamefont {Chen}}, \ and\ \bibinfo
  {author} {\bibfnamefont {Q.}~\bibnamefont {Shi}},\ }\href {\doibase
  10.1063/1.3604561} {\bibfield  {journal} {\bibinfo  {journal} {J. Chem.
  Phys.}\ }\textbf {\bibinfo {volume} {134}},\ \bibinfo {pages} {244116}
  (\bibinfo {year} {2011})}\BibitemShut {NoStop}%
\bibitem [{\citenamefont {Troisi}(2011)}]{Troisi_2011}%
  \BibitemOpen
  \bibfield  {author} {\bibinfo {author} {\bibfnamefont {A.}~\bibnamefont
  {Troisi}},\ }\href {\doibase 10.1039/C0CS00198H} {\bibfield  {journal}
  {\bibinfo  {journal} {Chem. Soc. Rev.}\ }\textbf {\bibinfo {volume} {40}},\
  \bibinfo {pages} {2347} (\bibinfo {year} {2011})}\BibitemShut {NoStop}%
\bibitem [{\citenamefont {Ciuchi}\ and\ \citenamefont
  {Fratini}(2012)}]{Fratini2012}%
  \BibitemOpen
  \bibfield  {author} {\bibinfo {author} {\bibfnamefont {S.}~\bibnamefont
  {Ciuchi}}\ and\ \bibinfo {author} {\bibfnamefont {S.}~\bibnamefont
  {Fratini}},\ }\href {\doibase 10.1103/PhysRevB.86.245201} {\bibfield
  {journal} {\bibinfo  {journal} {Phys. Rev. B}\ }\textbf {\bibinfo {volume}
  {86}},\ \bibinfo {pages} {245201} (\bibinfo {year} {2012})}\BibitemShut
  {NoStop}%
\bibitem [{\citenamefont {Fratini}, \citenamefont {Mayou},\ and\ \citenamefont
  {Ciuchi}(2016)}]{fratini_transient_2016}%
  \BibitemOpen
  \bibfield  {author} {\bibinfo {author} {\bibfnamefont {S.}~\bibnamefont
  {Fratini}}, \bibinfo {author} {\bibfnamefont {D.}~\bibnamefont {Mayou}}, \
  and\ \bibinfo {author} {\bibfnamefont {S.}~\bibnamefont {Ciuchi}},\ }\href
  {\doibase 10.1002/adfm.201502386} {\bibfield  {journal} {\bibinfo  {journal}
  {Adv. Funct. Mater.}\ }\textbf {\bibinfo {volume} {26}},\ \bibinfo {pages}
  {2292} (\bibinfo {year} {2016})}\BibitemShut {NoStop}%
\bibitem [{\citenamefont {Oberhofer}, \citenamefont {Reuter},\ and\
  \citenamefont {Blumberger}(2017)}]{oberhofer_charge_2017}%
  \BibitemOpen
  \bibfield  {author} {\bibinfo {author} {\bibfnamefont {H.}~\bibnamefont
  {Oberhofer}}, \bibinfo {author} {\bibfnamefont {K.}~\bibnamefont {Reuter}}, \
  and\ \bibinfo {author} {\bibfnamefont {J.}~\bibnamefont {Blumberger}},\
  }\href {\doibase 10.1021/acs.chemrev.7b00086} {\bibfield  {journal} {\bibinfo
   {journal} {Chem. Rev.}\ }\textbf {\bibinfo {volume} {117}},\ \bibinfo
  {pages} {10319} (\bibinfo {year} {2017})}\BibitemShut {NoStop}%
\bibitem [{\citenamefont {Giannini}\ \emph {et~al.}(2019)\citenamefont
  {Giannini}, \citenamefont {Carof}, \citenamefont {Ellis}, \citenamefont
  {Yang}, \citenamefont {Ziogos}, \citenamefont {Ghosh},\ and\ \citenamefont
  {Blumberger}}]{Blumberger2019}%
  \BibitemOpen
  \bibfield  {author} {\bibinfo {author} {\bibfnamefont {S.}~\bibnamefont
  {Giannini}}, \bibinfo {author} {\bibfnamefont {A.}~\bibnamefont {Carof}},
  \bibinfo {author} {\bibfnamefont {M.}~\bibnamefont {Ellis}}, \bibinfo
  {author} {\bibfnamefont {H.}~\bibnamefont {Yang}}, \bibinfo {author}
  {\bibfnamefont {O.~G.}\ \bibnamefont {Ziogos}}, \bibinfo {author}
  {\bibfnamefont {S.}~\bibnamefont {Ghosh}}, \ and\ \bibinfo {author}
  {\bibfnamefont {J.}~\bibnamefont {Blumberger}},\ }\href
  {https://doi.org/10.1038/s41467-019-11775-9} {\bibfield  {journal} {\bibinfo
  {journal} {Nat. Commun.}\ }\textbf {\bibinfo {volume} {10}},\ \bibinfo
  {pages} {3843} (\bibinfo {year} {2019})}\BibitemShut {NoStop}%
\bibitem [{\citenamefont {Asher}\ \emph {et~al.}(2020)\citenamefont {Asher},
  \citenamefont {Angerer}, \citenamefont {Korobko}, \citenamefont
  {Diskin-Posner}, \citenamefont {Egger},\ and\ \citenamefont
  {Yaffe}}]{Egger_Yaffe_2020}%
  \BibitemOpen
  \bibfield  {author} {\bibinfo {author} {\bibfnamefont {M.}~\bibnamefont
  {Asher}}, \bibinfo {author} {\bibfnamefont {D.}~\bibnamefont {Angerer}},
  \bibinfo {author} {\bibfnamefont {R.}~\bibnamefont {Korobko}}, \bibinfo
  {author} {\bibfnamefont {Y.}~\bibnamefont {Diskin-Posner}}, \bibinfo {author}
  {\bibfnamefont {D.~A.}\ \bibnamefont {Egger}}, \ and\ \bibinfo {author}
  {\bibfnamefont {O.}~\bibnamefont {Yaffe}},\ }\href {\doibase
  https://doi.org/10.1002/adma.201908028} {\bibfield  {journal} {\bibinfo
  {journal} {Adv. Mater.}\ }\textbf {\bibinfo {volume} {32}},\ \bibinfo {pages}
  {1908028} (\bibinfo {year} {2020})}\BibitemShut {NoStop}%
\bibitem [{\citenamefont {Giannini}\ \emph {et~al.}(2023)\citenamefont
  {Giannini}, \citenamefont {Di~Virgilio}, \citenamefont {Bardini},
  \citenamefont {Hausch}, \citenamefont {Geuchies}, \citenamefont {Zheng},
  \citenamefont {Volpi}, \citenamefont {Elsner}, \citenamefont {Broch},
  \citenamefont {Geerts}, \citenamefont {Schreiber}, \citenamefont
  {Schweicher}, \citenamefont {Wang}, \citenamefont {Blumberger}, \citenamefont
  {Bonn},\ and\ \citenamefont {Beljonne}}]{Beljonne_Blumberger_2023}%
  \BibitemOpen
  \bibfield  {author} {\bibinfo {author} {\bibfnamefont {S.}~\bibnamefont
  {Giannini}}, \bibinfo {author} {\bibfnamefont {L.}~\bibnamefont
  {Di~Virgilio}}, \bibinfo {author} {\bibfnamefont {M.}~\bibnamefont
  {Bardini}}, \bibinfo {author} {\bibfnamefont {J.}~\bibnamefont {Hausch}},
  \bibinfo {author} {\bibfnamefont {J.~J.}\ \bibnamefont {Geuchies}}, \bibinfo
  {author} {\bibfnamefont {W.}~\bibnamefont {Zheng}}, \bibinfo {author}
  {\bibfnamefont {M.}~\bibnamefont {Volpi}}, \bibinfo {author} {\bibfnamefont
  {J.}~\bibnamefont {Elsner}}, \bibinfo {author} {\bibfnamefont
  {K.}~\bibnamefont {Broch}}, \bibinfo {author} {\bibfnamefont {Y.~H.}\
  \bibnamefont {Geerts}}, \bibinfo {author} {\bibfnamefont {F.}~\bibnamefont
  {Schreiber}}, \bibinfo {author} {\bibfnamefont {G.}~\bibnamefont
  {Schweicher}}, \bibinfo {author} {\bibfnamefont {H.~I.}\ \bibnamefont
  {Wang}}, \bibinfo {author} {\bibfnamefont {J.}~\bibnamefont {Blumberger}},
  \bibinfo {author} {\bibfnamefont {M.}~\bibnamefont {Bonn}}, \ and\ \bibinfo
  {author} {\bibfnamefont {D.}~\bibnamefont {Beljonne}},\ }\href
  {https://doi.org/10.1038/s41563-023-01664-4} {\bibfield  {journal} {\bibinfo
  {journal} {Nat. Mater.}\ }\textbf {\bibinfo {volume} {22}},\ \bibinfo {pages}
  {1361} (\bibinfo {year} {2023})}\BibitemShut {NoStop}%
\bibitem [{\citenamefont {Lacroix}\ \emph {et~al.}(2020)\citenamefont
  {Lacroix}, \citenamefont {de~Laissardi\`ere}, \citenamefont {Qu\'emerais},
  \citenamefont {Julien},\ and\ \citenamefont {Mayou}}]{Lacroix2020}%
  \BibitemOpen
  \bibfield  {author} {\bibinfo {author} {\bibfnamefont {A.}~\bibnamefont
  {Lacroix}}, \bibinfo {author} {\bibfnamefont {G.~T.}\ \bibnamefont
  {de~Laissardi\`ere}}, \bibinfo {author} {\bibfnamefont {P.}~\bibnamefont
  {Qu\'emerais}}, \bibinfo {author} {\bibfnamefont {J.-P.}\ \bibnamefont
  {Julien}}, \ and\ \bibinfo {author} {\bibfnamefont {D.}~\bibnamefont
  {Mayou}},\ }\href {\doibase 10.1103/PhysRevLett.124.196601} {\bibfield
  {journal} {\bibinfo  {journal} {Phys. Rev. Lett.}\ }\textbf {\bibinfo
  {volume} {124}},\ \bibinfo {pages} {196601} (\bibinfo {year}
  {2020})}\BibitemShut {NoStop}%
\bibitem [{\citenamefont {Dörflinger}, \citenamefont {Rieder},\ and\
  \citenamefont {Dyakonov}(2025)}]{Dörflinger2025}%
  \BibitemOpen
  \bibfield  {author} {\bibinfo {author} {\bibfnamefont {P.}~\bibnamefont
  {Dörflinger}}, \bibinfo {author} {\bibfnamefont {P.}~\bibnamefont {Rieder}},
  \ and\ \bibinfo {author} {\bibfnamefont {V.}~\bibnamefont {Dyakonov}},\
  }\href {\doibase https://doi.org/10.1002/aenm.202403332} {\bibfield
  {journal} {\bibinfo  {journal} {Adv. Energy Mater.}\ ,\ \bibinfo {pages}
  {2403332}} (\bibinfo {year} {2025})}\BibitemShut {NoStop}%
\bibitem [{\citenamefont {Kresse}\ and\ \citenamefont
  {Furthmüller}(1996)}]{VASP}%
  \BibitemOpen
  \bibfield  {author} {\bibinfo {author} {\bibfnamefont {G.}~\bibnamefont
  {Kresse}}\ and\ \bibinfo {author} {\bibfnamefont {J.}~\bibnamefont
  {Furthmüller}},\ }\href {\doibase 10.1103/PhysRevB.54.11169} {\bibfield
  {journal} {\bibinfo  {journal} {Phys. Rev. B}\ }\textbf {\bibinfo {volume}
  {54}},\ \bibinfo {pages} {11169} (\bibinfo {year} {1996})}\BibitemShut
  {NoStop}%
\bibitem [{\citenamefont {Heyd}, \citenamefont {Scuseria},\ and\ \citenamefont
  {Ernzerhof}(2003)}]{HSE}%
  \BibitemOpen
  \bibfield  {author} {\bibinfo {author} {\bibfnamefont {J.}~\bibnamefont
  {Heyd}}, \bibinfo {author} {\bibfnamefont {G.~E.}\ \bibnamefont {Scuseria}},
  \ and\ \bibinfo {author} {\bibfnamefont {M.}~\bibnamefont {Ernzerhof}},\
  }\href {\doibase 10.1063/1.1564060} {\bibfield  {journal} {\bibinfo
  {journal} {Chem. Phys.}\ }\textbf {\bibinfo {volume} {118}},\ \bibinfo
  {pages} {8207} (\bibinfo {year} {2003})}\BibitemShut {NoStop}%
\bibitem [{\citenamefont {Heyd}, \citenamefont {Scuseria},\ and\ \citenamefont
  {Ernzerhof}(2006)}]{Erratum_HSE}%
  \BibitemOpen
  \bibfield  {author} {\bibinfo {author} {\bibfnamefont {J.}~\bibnamefont
  {Heyd}}, \bibinfo {author} {\bibfnamefont {G.~E.}\ \bibnamefont {Scuseria}},
  \ and\ \bibinfo {author} {\bibfnamefont {M.}~\bibnamefont {Ernzerhof}},\
  }\href {\doibase 10.1063/1.2204597} {\bibfield  {journal} {\bibinfo
  {journal} {Chem. Phys.}\ }\textbf {\bibinfo {volume} {124}},\ \bibinfo
  {pages} {219906} (\bibinfo {year} {2006})}\BibitemShut {NoStop}%
\bibitem [{\citenamefont {Perdew}, \citenamefont {Burke},\ and\ \citenamefont
  {Ernzerhof}(1996)}]{PBE}%
  \BibitemOpen
  \bibfield  {author} {\bibinfo {author} {\bibfnamefont {J.~P.}\ \bibnamefont
  {Perdew}}, \bibinfo {author} {\bibfnamefont {K.}~\bibnamefont {Burke}}, \
  and\ \bibinfo {author} {\bibfnamefont {M.}~\bibnamefont {Ernzerhof}},\ }\href
  {\doibase 10.1103/PhysRevLett.77.3865} {\bibfield  {journal} {\bibinfo
  {journal} {Phys. Rev. Lett.}\ }\textbf {\bibinfo {volume} {77}},\ \bibinfo
  {pages} {3865} (\bibinfo {year} {1996})}\BibitemShut {NoStop}%
\bibitem [{\citenamefont {Yuan}\ \emph {et~al.}(2015)\citenamefont {Yuan},
  \citenamefont {Xu}, \citenamefont {Xu}, \citenamefont {Hong}, \citenamefont
  {Xu},\ and\ \citenamefont {Wang}}]{bandgap_2015}%
  \BibitemOpen
  \bibfield  {author} {\bibinfo {author} {\bibfnamefont {Y.}~\bibnamefont
  {Yuan}}, \bibinfo {author} {\bibfnamefont {R.}~\bibnamefont {Xu}}, \bibinfo
  {author} {\bibfnamefont {H.-T.}\ \bibnamefont {Xu}}, \bibinfo {author}
  {\bibfnamefont {F.}~\bibnamefont {Hong}}, \bibinfo {author} {\bibfnamefont
  {F.}~\bibnamefont {Xu}}, \ and\ \bibinfo {author} {\bibfnamefont {L.-J.}\
  \bibnamefont {Wang}},\ }\href {\doibase 10.1088/1674-1056/24/11/116302}
  {\bibfield  {journal} {\bibinfo  {journal} {Chin. Phys. B}\ }\textbf
  {\bibinfo {volume} {24}},\ \bibinfo {pages} {116302} (\bibinfo {year}
  {2015})}\BibitemShut {NoStop}%
\bibitem [{\citenamefont {Das}, \citenamefont {Di~Liberto},\ and\ \citenamefont
  {Pacchioni}(2022)}]{bandgap_PBE_HSE}%
  \BibitemOpen
  \bibfield  {author} {\bibinfo {author} {\bibfnamefont {T.}~\bibnamefont
  {Das}}, \bibinfo {author} {\bibfnamefont {G.}~\bibnamefont {Di~Liberto}}, \
  and\ \bibinfo {author} {\bibfnamefont {G.}~\bibnamefont {Pacchioni}},\ }\href
  {\doibase 10.1021/acs.jpcc.1c09594} {\bibfield  {journal} {\bibinfo
  {journal} {J. Phys. Chem. C}\ }\textbf {\bibinfo {volume} {126}},\ \bibinfo
  {pages} {2184} (\bibinfo {year} {2022})}\BibitemShut {NoStop}%
\bibitem [{\citenamefont {Pizzi}\ \emph {et~al.}(2020)\citenamefont {Pizzi},
  \citenamefont {Vitale}, \citenamefont {Arita}, \citenamefont {Blügel},
  \citenamefont {Freimuth}, \citenamefont {Géranton}, \citenamefont
  {Gibertini}, \citenamefont {Gresch}, \citenamefont {Johnson}, \citenamefont
  {Koretsune}, \citenamefont {Ibañez-Azpiroz}, \citenamefont {Lee},
  \citenamefont {Lihm}, \citenamefont {Marchand}, \citenamefont {Marrazzo},
  \citenamefont {Mokrousov}, \citenamefont {Mustafa}, \citenamefont {Nohara},
  \citenamefont {Nomura}, \citenamefont {Paulatto}, \citenamefont {Poncé},
  \citenamefont {Ponweiser}, \citenamefont {Qiao}, \citenamefont {Thöle},
  \citenamefont {Tsirkin}, \citenamefont {Wierzbowska}, \citenamefont
  {Marzari}, \citenamefont {Vanderbilt}, \citenamefont {Souza}, \citenamefont
  {Mostofi},\ and\ \citenamefont {Yates}}]{W90_2}%
  \BibitemOpen
  \bibfield  {author} {\bibinfo {author} {\bibfnamefont {G.}~\bibnamefont
  {Pizzi}}, \bibinfo {author} {\bibfnamefont {V.}~\bibnamefont {Vitale}},
  \bibinfo {author} {\bibfnamefont {R.}~\bibnamefont {Arita}}, \bibinfo
  {author} {\bibfnamefont {S.}~\bibnamefont {Blügel}}, \bibinfo {author}
  {\bibfnamefont {F.}~\bibnamefont {Freimuth}}, \bibinfo {author}
  {\bibfnamefont {G.}~\bibnamefont {Géranton}}, \bibinfo {author}
  {\bibfnamefont {M.}~\bibnamefont {Gibertini}}, \bibinfo {author}
  {\bibfnamefont {D.}~\bibnamefont {Gresch}}, \bibinfo {author} {\bibfnamefont
  {C.}~\bibnamefont {Johnson}}, \bibinfo {author} {\bibfnamefont
  {T.}~\bibnamefont {Koretsune}}, \bibinfo {author} {\bibfnamefont
  {J.}~\bibnamefont {Ibañez-Azpiroz}}, \bibinfo {author} {\bibfnamefont
  {H.}~\bibnamefont {Lee}}, \bibinfo {author} {\bibfnamefont {J.-M.}\
  \bibnamefont {Lihm}}, \bibinfo {author} {\bibfnamefont {D.}~\bibnamefont
  {Marchand}}, \bibinfo {author} {\bibfnamefont {A.}~\bibnamefont {Marrazzo}},
  \bibinfo {author} {\bibfnamefont {Y.}~\bibnamefont {Mokrousov}}, \bibinfo
  {author} {\bibfnamefont {J.~I.}\ \bibnamefont {Mustafa}}, \bibinfo {author}
  {\bibfnamefont {Y.}~\bibnamefont {Nohara}}, \bibinfo {author} {\bibfnamefont
  {Y.}~\bibnamefont {Nomura}}, \bibinfo {author} {\bibfnamefont
  {L.}~\bibnamefont {Paulatto}}, \bibinfo {author} {\bibfnamefont
  {S.}~\bibnamefont {Poncé}}, \bibinfo {author} {\bibfnamefont
  {T.}~\bibnamefont {Ponweiser}}, \bibinfo {author} {\bibfnamefont
  {J.}~\bibnamefont {Qiao}}, \bibinfo {author} {\bibfnamefont {F.}~\bibnamefont
  {Thöle}}, \bibinfo {author} {\bibfnamefont {S.~S.}\ \bibnamefont {Tsirkin}},
  \bibinfo {author} {\bibfnamefont {M.}~\bibnamefont {Wierzbowska}}, \bibinfo
  {author} {\bibfnamefont {N.}~\bibnamefont {Marzari}}, \bibinfo {author}
  {\bibfnamefont {D.}~\bibnamefont {Vanderbilt}}, \bibinfo {author}
  {\bibfnamefont {I.}~\bibnamefont {Souza}}, \bibinfo {author} {\bibfnamefont
  {A.~A.}\ \bibnamefont {Mostofi}}, \ and\ \bibinfo {author} {\bibfnamefont
  {J.~R.}\ \bibnamefont {Yates}},\ }\href {\doibase 10.1088/1361-648X/ab51ff}
  {\bibfield  {journal} {\bibinfo  {journal} {J. Phys.: Condens. Matter}\
  }\textbf {\bibinfo {volume} {32}},\ \bibinfo {pages} {165902} (\bibinfo
  {year} {2020})}\BibitemShut {NoStop}%
\bibitem [{Note1()}]{Note1}%
  \BibitemOpen
  \bibinfo {note} {See Supplemental Material at LINK for more details on the
  impact of the functional on the hopping elements and the connection of the
  Crystal Orbital Bond Index to the hopping elements.}\BibitemShut {Stop}%
\bibitem [{\citenamefont {Ashhab}\ \emph {et~al.}(2017)\citenamefont {Ashhab},
  \citenamefont {Voznyy}, \citenamefont {Hoogland}, \citenamefont {Sargent},\
  and\ \citenamefont {Madjet}}]{Ashhab2017}%
  \BibitemOpen
  \bibfield  {author} {\bibinfo {author} {\bibfnamefont {S.}~\bibnamefont
  {Ashhab}}, \bibinfo {author} {\bibfnamefont {O.}~\bibnamefont {Voznyy}},
  \bibinfo {author} {\bibfnamefont {S.}~\bibnamefont {Hoogland}}, \bibinfo
  {author} {\bibfnamefont {E.~H.}\ \bibnamefont {Sargent}}, \ and\ \bibinfo
  {author} {\bibfnamefont {M.~E.}\ \bibnamefont {Madjet}},\ }\href {\doibase
  10.1038/s41598-017-09442-4} {\bibfield  {journal} {\bibinfo  {journal} {Sci.
  Rep.}\ }\textbf {\bibinfo {volume} {7}},\ \bibinfo {pages} {8902} (\bibinfo
  {year} {2017})}\BibitemShut {NoStop}%
\bibitem [{\citenamefont {Mehdizadeh}, \citenamefont {Akhtarianfar},\ and\
  \citenamefont {Shojaei}(2019)}]{Mehdizadeh2019}%
  \BibitemOpen
  \bibfield  {author} {\bibinfo {author} {\bibfnamefont {A.}~\bibnamefont
  {Mehdizadeh}}, \bibinfo {author} {\bibfnamefont {S.~F.}\ \bibnamefont
  {Akhtarianfar}}, \ and\ \bibinfo {author} {\bibfnamefont {S.}~\bibnamefont
  {Shojaei}},\ }\href {\doibase 10.1021/acs.jpcc.8b11422} {\bibfield  {journal}
  {\bibinfo  {journal} {J. Phys. Chem. C}\ }\textbf {\bibinfo {volume} {123}},\
  \bibinfo {pages} {6725} (\bibinfo {year} {2019})}\BibitemShut {NoStop}%
\bibitem [{\citenamefont {Atkins}\ and\ \citenamefont
  {Friedman}(2005)}]{Atkins_Friedman}%
  \BibitemOpen
  \bibfield  {author} {\bibinfo {author} {\bibfnamefont {P.~W.}\ \bibnamefont
  {Atkins}}\ and\ \bibinfo {author} {\bibfnamefont {R.~S.}\ \bibnamefont
  {Friedman}},\ }\href@noop {} {\emph {\bibinfo {title} {Molecular Quantum
  Mechanics}}}\ (\bibinfo  {publisher} {Oxford University Press},\ \bibinfo
  {year} {2005})\BibitemShut {NoStop}%
\bibitem [{\citenamefont {Fetherolf}, \citenamefont
  {Gole\ifmmode~\check{z}\else \v{z}\fi{}},\ and\ \citenamefont
  {Berkelbach}(2020)}]{Berkelbach2020}%
  \BibitemOpen
  \bibfield  {author} {\bibinfo {author} {\bibfnamefont {J.~H.}\ \bibnamefont
  {Fetherolf}}, \bibinfo {author} {\bibfnamefont {D.}~\bibnamefont
  {Gole\ifmmode~\check{z}\else \v{z}\fi{}}}, \ and\ \bibinfo {author}
  {\bibfnamefont {T.~C.}\ \bibnamefont {Berkelbach}},\ }\href {\doibase
  10.1103/PhysRevX.10.021062} {\bibfield  {journal} {\bibinfo  {journal} {Phys.
  Rev. X}\ }\textbf {\bibinfo {volume} {10}},\ \bibinfo {pages} {021062}
  (\bibinfo {year} {2020})}\BibitemShut {NoStop}%
\bibitem [{\citenamefont {Paschal}, \citenamefont {Pogrebnoi},\ and\
  \citenamefont {Pogrebnaya}(2021)}]{Paschal2021}%
  \BibitemOpen
  \bibfield  {author} {\bibinfo {author} {\bibfnamefont {C.}~\bibnamefont
  {Paschal}}, \bibinfo {author} {\bibfnamefont {A.}~\bibnamefont {Pogrebnoi}},
  \ and\ \bibinfo {author} {\bibfnamefont {T.}~\bibnamefont {Pogrebnaya}},\
  }\href {\doibase 10.1007/s00339-021-04504-x} {\bibfield  {journal} {\bibinfo
  {journal} {Appl. Phys. A}\ }\textbf {\bibinfo {volume} {127}},\ \bibinfo
  {pages} {355} (\bibinfo {year} {2021})}\BibitemShut {NoStop}%
\bibitem [{\citenamefont {Herz}(2017)}]{Herz2017}%
  \BibitemOpen
  \bibfield  {author} {\bibinfo {author} {\bibfnamefont {L.~M.}\ \bibnamefont
  {Herz}},\ }\href {\doibase 10.1021/acsenergylett.7b00276} {\bibfield
  {journal} {\bibinfo  {journal} {ACS Energy Lett.}\ }\textbf {\bibinfo
  {volume} {2}},\ \bibinfo {pages} {1539} (\bibinfo {year} {2017})}\BibitemShut
  {NoStop}%
\bibitem [{\citenamefont {Capitaine}\ and\ \citenamefont
  {Sciacca}(2021)}]{Capitaine2021}%
  \BibitemOpen
  \bibfield  {author} {\bibinfo {author} {\bibfnamefont {A.}~\bibnamefont
  {Capitaine}}\ and\ \bibinfo {author} {\bibfnamefont {B.}~\bibnamefont
  {Sciacca}},\ }\href {\doibase https://doi.org/10.1002/adma.202102588}
  {\bibfield  {journal} {\bibinfo  {journal} {Adv. Mater.}\ }\textbf {\bibinfo
  {volume} {33}},\ \bibinfo {pages} {2102588} (\bibinfo {year}
  {2021})}\BibitemShut {NoStop}%
\end{thebibliography}%

\end{document}